\documentclass[pre,preprint,preprintnumbers,amsmath,amssymb,showpacs]{revtex4}
\usepackage{graphicx,bm,amsmath,epsfig}

\begin{document}


\def\reff#1{(\ref{#1})}
\newcommand{\be}{\begin{equation}}
\newcommand{\ee}{\end{equation}}
\newcommand{\<}{\langle}
\renewcommand{\>}{\rangle}
\newcommand{\eps}{\epsilon}

\def\spose#1{\hbox to 0pt{#1\hss}}
\def\ltapprox{\mathrel{\spose{\lower 3pt\hbox{$\mathchar"218$}}
 \raise 2.0pt\hbox{$\mathchar"13C$}}}
\def\gtapprox{\mathrel{\spose{\lower 3pt\hbox{$\mathchar"218$}}
 \raise 2.0pt\hbox{$\mathchar"13E$}}}

\def\bsigma{\mbox{\protect\boldmath $\sigma$}}
\def\bpi{\mbox{\protect\boldmath $\pi$}}
\def\smfrac#1#2{{\textstyle\frac{#1}{#2}}}
\def\smhalf{ {\smfrac{1}{2}} }

\newcommand{\re}{\mathop{\rm Re}\nolimits}
\newcommand{\im}{\mathop{\rm Im}\nolimits}
\newcommand{\tr}{\mathop{\rm tr}\nolimits}
\newcommand{\fr}{\frac}

\def\Z{{\mathbb Z}}
\def\R{{\mathbb R}}
\def\C{{\mathbb C}}

\title{Colloids and polymers in random colloidal matrices: \\
demixing under good-solvent conditions}

 \author{Mario Alberto Annunziata\footnote{Present address: 
Institut f\"ur Theoretische Physik II, 
Heinrich-Heine-Universit\"at D\"usseldorf,
Universit\"atsstrasse 1, D-40225 D\"usseldorf, Germany} }
 \affiliation{CNR, Istituto dei Sistemi Complessi \\
         (Area della Ricerca di Roma Tor Vergata) \\
          Via del Fosso del Cavaliere 100, I-00133 Roma, Italy \\
   e-mail: {\tt m.annunziata@isc.cnr.it}}
 \author{ Andrea Pelissetto }
 \affiliation{
   Dipartimento di Fisica, Universit\`a degli Studi di Roma ``La Sapienza'' \\
   and INFN -- Sezione di Roma I  \\
   Piazzale A. Moro 2, I-00185 Roma, Italy \\
  e-mail: {\tt Andrea.Pelissetto@roma1.infn.it }}


\begin{abstract}
We consider a simplified coarse-grained model for colloid-polymer 
mixtures, in which polymers are represented as monoatomic molecules
interacting by means of pair potentials. We use it to study 
polymer-colloid segregation in the presence of a quenched 
matrix of colloidal hard spheres. We fix the polymer-to-colloid size
ratio to 0.8 and consider matrices such that the fraction $f$ of the 
volume that is not accessible to the colloids due to the matrix is 
equal to 40\%. As in the Asakura-Oosawa-Vrij (AOV) case, we find that 
binodal curves in the polymer and colloid volume-fraction plane 
have a small dependence on disorder. As for the position of the 
critical point, the behavior is different from that observed in the 
AOV case: while the critical colloid volume fraction is essentially
the same in the bulk and in the presence of the matrix, the 
polymer volume fraction at criticality increases as $f$ increases.
At variance with the AOV case, no capillary colloid condensation or evaporation
is generically observed.
\end{abstract}


\maketitle
\thispagestyle{empty}   

\clearpage

\section{Introduction}

The study of the properties of fluids in porous materials \cite{GGRS-99}
is very important 
for its technological applications in many fields of science, from physics
and chemistry to geology, engineering, and even agriculture. In this paper we 
consider the demixing of a binary mixture of nonadsorbing colloids of radius 
$R_c$ and of polymers of radius of gyration $R_g$, with the 
purpose of understanding how the porous structure influences the phase 
behavior. We are considering mesoporous disordered materials, like 
silica gels, which present large pores and can thus adsorb mesoscopic 
particles like colloids. 
In the bulk the phase behavior of colloid-polymer mixtures
depends on the parameter $q = R_g/R_c$. For $q$ small, 
$q < q_c$ ($q_c\approx 0.2$-0.3 under good-solvent conditions), 
only a fluid-solid transition is observed, 
analogous to that observed in hard-sphere systems. For $q > q_c$
also a fluid-fluid transition occurs, between a colloid-liquid (polymer-poor)
and a colloid-gas (polymer-rich) phase. It is such a fluid-fluid transition 
that will be investigated in the present paper.

At least qualitatively, many aspects of the behavior of polymer-colloid mixtures
can be understood by using the  Asakura-Oosawa-Vrij (AOV) model
\cite{AO-54,Vrij-76}, which gives a coarse-grained (CG) description of the
mixture. Polymers are treated as an ideal gas
of point particles, whose radius is identified with the
radius of gyration $R_g$, which interact with the colloids (hard spheres of 
radius $R_c$) by means of a simple hard-core potential.
This model is extremely crude since it ignores the polymeric structure and
polymer-polymer repulsion, which is relevant in the good-solvent regime.
Nonetheless, it correctly predicts
polymer-colloid demixing as a result of the entropy-driven
effective attraction (depletion interaction)
between colloidal pairs due to the presence of the polymers
\cite{MF-94,DBE-99,VH-04bis,VHB-05,BLH-02,DvR-02,GHR-83,LPPSW-92,SLBE-00}.
AOV colloid-polymer mixtures in a porous matrix have been
studied in Refs.~\cite{SSKK-02,VBL-06,VBL-08,PVCL-08,Vink-09,AP-11} by means
of density-functional theory, integral equations, and
Monte Carlo (MC) simulations. The nature of the
critical transition has been fully clarified
\cite{VBL-06,VBL-08,PVCL-08,Vink-09}:
if the obstacles are random and there is a preferred affinity
of the quenched obstacles to one of the phases,
the transition is in the same universality class as that
occurring in the random-field Ising model,
in agreement with a general argument by de Gennes \cite{deGennes-84}.
If these conditions are not satisfied, standard Ising or randomly
dilute Ising behavior is observed instead, see Refs.~\cite{DSLP-08,FV-11}.
Recently, we considered the AOV model and 
investigated \cite{AP-11} how demixing is
influenced by the amount of disorder and by its nature. 
We found that demixing was, to a large extent, only dependent on 
the fraction $f$ of the volume that is not accessible to the colloids due to 
the presence of the random matrix. The matrix topology was instead 
largely irrelevant.
Moreover, we observed the possibility of capillary condensation of the colloids:
for some values of the parameters, a colloid-gas bulk phase is in
equilibrium with a colloid-liquid phase in the matrix. 

The AOV model completely neglects polymer-polymer interactions, hence it may
only be quantitatively predictive close to the $\theta$ point where,
to some extent, polymers behave as ideal particles. 
In this paper we make a first step towards the inclusion of 
polymer-polymer interactions, allowing us to study systems under
good-solvent conditions. 
We still use a CG model in which 
polymers are treated as monoatomic molecules, but we include a polymer-polymer
repulsive 
pair potential which is such to reproduce the correct thermodynamics in 
the low-density limit. In recent years, this class of CG models has 
been extensively studied \cite{Likos-01,LBHM-00,BLHM-01}.
It is now clear that, unless one includes 
many-body interactions or considers density-dependent potentials, they
are quantitatively predictive only in the dilute regime in which 
the polymer volume fraction $\eta_p$ [$\eta_p = c_p/c^*_p$, 
where $c_p = N_p/V$ is the concentration and 
$c^*_p = 3/(4 \pi R_g^3)$] satisfies $\eta_p\lesssim 1$.
If $q \lesssim 1$ (the so-called colloid regime),
fluid-fluid demixing occurs for $\eta_p \lesssim 1$,
hence monoatomic CG models are expected to provide reasonably accurate results.
If, instead, $q \gtrsim 1$ (the protein regime), demixing occurs for 
$\eta_p \gtrsim 1$, hence one must use more sophisticated CG multiblob
models \cite{multiblob}
in which each polymer is represented by a polyatomic molecule and
each atom corresponds to a polymer blob of size of the order of that 
of the colloid. In this paper, we investigate the case $q = 0.8$,
as in our previous work \cite{AP-11}. Since colloids are larger than polymers,
a CG approach based on a single-blob representation of the 
polymers should be adequate.

The exact polymer-polymer and polymer-colloid pair potentials appropriate 
for single-blob models have been computed in several papers
\cite{DH-94,BLHM-01,BLH-02,BL-02,PH-05,PH-06}. The polymer-polymer
potential has a Gaussian shape with a width of the order of $R_g$ 
and is significantly different from zero ($u_{pp}(r)/u_{pp}(0) \gtrsim 0.01$)
up to $r \approx 2.5 R_g$. Analogously, the typical colloid-polymer 
potential has a tail, which is still significant for $r\approx 2(R_g + R_c)$. 
The presence of these
tails makes simulations quite slow, since one must consider a large 
number of neighboring molecules in each updating step. 
Since our simulations are extremely complex and time-consuming---we
must average over a large number of disorder realizations---we decided to 
replace the exact potentials with simpler ones 
that generalize the AOV interactions. They are square potentials, hence have
no tails, and thus allow us to determinate the energy quite fast. Of course,
this simplification implies that our results cannot be 
quantitatively accurate.
Still, our simulations should allow us to understand how the presence of 
the porous matrix changes the behavior of the polymer-colloid mixture 
under good-solvent conditions.

The paper is organized as follows. In Sec.~\ref{sec2} 
we present the simple model which we will use and give the basic definitions.
In Sec.~\ref{sec3} we discuss the model in the bulk, while in 
Sec.~\ref{sec4} we determine the binodal lines and the critical-point
positions for polymers and colloids adsorbed in two different 
colloidal matrices.  Finally, in Sec.~\ref{sec5} we present our conclusions. 

\section{Definitions} \label{sec2}

\subsection{Models}

In this paper we model polymers as soft effective spheres.
Polymers of radius of gyration $R_g$ are represented by 
monoatomic molecules interacting with pair potential
\begin{eqnarray}
u_{pp}(r) &=& \begin{cases}
   \epsilon_{pp} & \text{$\qquad$ for $r < \alpha R_g$,} \cr 
   0 & \text{$\qquad$ for $r \ge \alpha R_g$,}
   \end{cases} 
\label{upp}
\end{eqnarray}
where $\alpha$ and $\epsilon_{pp}$ are parameters 
which are fixed below. Since we expect demixing to occur for 
$\eta_p\lesssim 1$, we have fixed $\alpha$ and $\epsilon_{pp}$ 
to reproduce accurately the compressibility factor 
$Z = p/(k_B T c)$ ($p$ is the pressure and $c$ the concentration)
in this density interval. In practice, we have determined the 
second virial coefficient $B_{2,pp}$ and $Z$ at $\eta_p=1$ using 
model (\ref{upp}) for several values of the parameters
and we have compared the results with 
those obtained in full-monomer simulations
\cite{CMP-06,Pelissetto-08}. Requiring the 
model to reproduce the estimate \cite{CMP-06} 
$A_{2,pp} = B_{2,pp} R_g^{-3} \approx 5.50$
and the estimate \cite{Pelissetto-08} of $Z$ for $\eta_p=1$, 
we obtain
\begin{equation}
\alpha = 1.58\qquad \qquad \epsilon_{pp} = 1.096.
\end{equation}
It is important to note that, although the thermodynamics is quite well
reproduced up to $\eta_p \approx 1$ and with small errors up to 
$\eta_p = 2$ (at such value of $\eta_p$ the difference between $Z$ computed in 
the present model and that for polymers is 8\%), the intermolecular 
structure is poorly reproduced. For instance, in this model the 
intermolecular distribution function is discontinuous and oscillates 
due to the discontinuity of the potential, a behavior which is not
observed in polymer systems.

Let us now introduce the colloids. Two colloids interact with a 
hard-sphere potential 
\begin{eqnarray}
u_{cc}(r) &=& \begin{cases}
   \infty & \text{$\qquad$ for $r < 2 R_c$,} \cr 
   0 & \text{$\qquad$ for $r \ge 2 R_c$,}
   \end{cases} 
\end{eqnarray}
while colloid-polymer interactions are modelled by taking a soft version
of the AOV hard-core potential
\begin{eqnarray}
u_{cp}(r) &=& \begin{cases}
   \epsilon_{cp}(q) & \text{$\qquad$ for $r < R_c + R_g$,} \cr 
   0 & \text{$\qquad$ for $r \ge R_c + R_g$.}
   \end{cases} 
\label{ucp-def}
\end{eqnarray}
The parameter $\epsilon_{cp}(q)$ is fixed so that the thermodynamics is exactly
reproduced in the low-density limit. For this purpose we compute 
the universal polymer-colloid second-virial combination
\begin{eqnarray}
A_{2,cp} &=& 
B_{2,cp} R_c^{-3/2} R_g^{-3/2} = 
    {1\over2} R_c^{-3/2} R_g^{-3/2} 
    \int d^3{\bf r} \left(1 - e^{-u_{cp}(r)}\right) = 
\nonumber \\
  &=& {2\pi\over3} (1 - e^{-\epsilon_{cp}}) (1 + q)^3 q^{-3/2}.
\end{eqnarray}
Estimates of 
$A_{2,cp}$ were obtained in Ref. \cite{PH-06} for polymers under good-solvent 
conditions. We fix $\epsilon_{cp}(q)$ so that the value of $A_{2,cp}$  
in our model is the same as that obtained in Ref.~\cite{PH-06}.
Results are reported in Table~\ref{table-epscp}.
As expected, $\epsilon_{cp}(q)$ diverges as $q\to 0$, since in this limit
polymers are quite well described by hard spheres.
In the opposite limit instead, the interaction energy vanishes, 
a phenomenon which is related to the fact that, for large values of $q$,
the polymer can wrap around the hard spheres. Hence, in the CG model 
in which each polymer is replaced by a monoatomic molecule positioned in the 
polymer center of mass, 
there is a significant probability that the CG polymer and the colloid are
one on top of the other. If we compare potentials (\ref{ucp-def}) 
with the exact ones reported in Refs.~\cite{BL-02,PH-06}, we observe that 
$u_{cp}(r)$, which  is essentially an average potential, overestimates 
the interaction energy for $R_g \lesssim r \lesssim R_c + R_g$, 
while it significantly underestimates $u_{cp}(r)$ close to overlap 
(for instance, the correct $u_{cp}(r)$ diverges for $r\to 0$ if $q\le 1$).
Note that a more accurate model could have been obtained by also introducing
a parameter $\alpha_{cp}$ to specify the range of the polymer-colloid
interactions, as in Eq.~(\ref{upp}). However, to fix an additional parameter 
we would have needed some additional thermodynamic information (for instance,
the pressure for a finite value of the polymer and colloid densities), which 
was not available to us. 

\begin{table}
\begin{tabular}{ccc}
\hline\hline
$q$  &   $\epsilon_{cp}$ \\
\hline
5      &    0.346 \\
2.5    &    0.725 \\
1.25   &    1.363  \\
1      &    1.642  \\
0.8    &    2.035  \\
\hline\hline
\end{tabular}
\caption{Estimates of the parameter $\epsilon_{cp}$.}
\label{table-epscp}
\end{table}

We mention that a different simplified model including polymer-polymer
interactions was introduced in Ref.~\cite{ZVBHV-09}
and studied for $q=0.8$. However,
with their parameter choices, thermodynamics is not reproduced in 
the low-density limit. Indeed, if we consider the second virial coefficients,
their model gives $A_{2,pp} \approx 1.79$ and $A_{2,cp} \approx 17.2$
to be compared with estimates $A_{2,pp} \approx 5.50$ and 
$A_{2,cp} \approx 14.8$ obtained from direct polymer simulations
\cite{CMP-06,PH-06}. While the colloid-polymer coefficient 
is reasonably close to the correct one (they differ by 15\%), 
the second virial coefficient for polymers is significantly smaller.
Hence, the model of Ref.~\cite{ZVBHV-09} appears to be more appropriate to 
describe polymers in the crossover region close to the 
$\theta$ point than in the good-solvent regime.

The colloidal matrix has been introduced as in 
our previous work. We consider a random distribution 
of quenched hard spheres of radius $R_{\rm dis}$. The
matrix-colloid and matrix-polymer interaction potentials are therefore:
\begin{eqnarray}
u_{cd}(r) &=& \begin{cases}
   \infty & \text{$\qquad$ for $r < R_{\rm dis} + R_c$,} \cr 
   0 & \text{$\qquad$ for $r \ge R_{\rm dis} + R_c$,}
   \end{cases} \nonumber \\[2mm]
u_{pd}(r) &=& \begin{cases}
   \epsilon_{cp}(q_{\rm dis}) & \text{$\qquad$ for $r < R_{\rm dis} + R_g$,} 
    \cr 
   0 & \text{$\qquad$ for $r \ge R_{\rm dis} + R_g$,}
   \end{cases} 
\end{eqnarray}
where $q_{\rm dis} \equiv R_g/R_{\rm dis}$. 
For $R_{\rm dis}\to 0$,
i.e., for $q_{\rm dis}\to \infty$, we have $\epsilon_{cq}(q_{\rm dis})\to 0$. 
As we already discussed,
this  reflects the fact that the polymer
can easily wrap around the small quenched colloid, which implies that the CG
polymer can easily overlap with it. As a consequence of this,
the matrix becomes less and less repulsive as $q_{\rm dis}$ increases, 
so that in the limit $R_{\rm dis} \to 0$, i.e. $q_{\rm dis}\to \infty$,  
we obtain bulk behavior.
This is, of course, an artifact of the model, which can only be eliminated
by using multiblob approaches \cite{multiblob}, 
in which a single polymer is represented
by many blobs, whose size is of the order of that of the quenched colloid.

In the simple model we consider, disorder is characterized by two parameters,
the reduced concentration
$\hat{c} \equiv c_{\rm dis} R_c^3$ ($c_{\rm dis} = N_{\rm dis}/V$,
where $N_{\rm dis}$ is the number of quenched hard spheres present in the 
volume $V$) and the ratio $R_{\rm dis}/R_c$.
However, as we already discussed in Ref.~\cite{AP-11}, it is much
more useful to characterize the amount of disorder by using the 
volume fraction $f$ which is not accessible to the colloids due to 
the matrix. To define it precisely, consider the region ${\cal R}$
in which the (centers of the) colloids are allowed:
\begin{equation}
{\cal R} = \{ {\bf r}: |{\bf r} - {\bf r}_i| \ge R_c + R_{\rm dis} ,
\hbox{
for all $1\le i\le N_{\rm dis}$ }
\},
\end{equation}
where ${\bf r}_i$ is the position of the $i$-th hard sphere belonging
to the matrix. If $V_{\cal R}$ is the volume of the region ${\cal R}$,
we define
\begin{equation}
   f \equiv 1 - {[V_{\cal R}]\over V},
\end{equation}
where $[V_{\cal R}]$ is the average of $V_{\cal R}$ over the different matrix
realizations.

\subsection{Simulation details} \label{sec2.2}

In this work we investigate the effect of disorder on the
fluid-fluid binodals for $q = 0.8$ which is the case we investigated in 
our previous work \cite{AP-11}.  We perform
simulations in the absence of the porous matrix and 
for $f=0.4$,  $R_{\rm dis}/R_c = 0.2,1.0$. 
In order to determine the coexistence curves we 
combine the grand-canonical algorithm with
the umbrella sampling \cite{TV-77} and
the simulated-tempering method \cite{MP-92},
as discussed in the Appendix of Ref.~\cite{AP-11}.

The grand partition sum for each disorder realization is
\begin{equation}
\Xi(V,z_p,z_c) = \sum_{N_p,N_c} z_p^{N_p} z_c^{N_c} Q(V,N_p,N_c),
\label{def-Xi}
\end{equation}
where $Q(V,N_p,N_c)$ is the configurational partition function of
a system of $N_p$ polymers and $N_c$ colloids in a volume $V$,
and $z_p$ and $z_c$ are the corresponding fugacities.
In Eq.~(\ref{def-Xi}) we normalize $Q(V,N_p,N_c)$ so that
$Q(V,1,0) = Q(V,0,1) = V$, hence $z_p$ and $z_c$ are dimensionful
parameters. 

The system shows, both in the bulk and in the matrix, a demixing transition.
For $z_p < z_{p,\rm crit}$ a single phase exists, while for 
$z_p > z_{p,\rm crit}$ coexistence is observed along the 
line $z_c = z_c^*(z_p)$.
In the MC simulations the position of the demixing curve
can be determined by studying the disorder averaged histograms of $N_c$
and $N_p$, which are defined as
\begin{eqnarray}
h_{c,\rm ave} (N_{c,0},z_p,z_c) &\equiv &
  \left[ \left\< \delta(N_c,N_{c,0})\right\>_{GC,z_p,z_c}\right] ,
\nonumber \\
h_{p,\rm ave} (N_{p,0},z_p,z_c) &\equiv &
  \left[ \left\< \delta(N_p,N_{p,0})\right\>_{GC,z_p,z_c}\right] ,
\label{hphc-def}
\end{eqnarray}
where $\delta({x,y})$ is Kronecker's delta [$\delta({x,x}) = 1$,
$\delta({x,y}) = 0$ for $x\not= y$], $\left\<\cdot \right\>_{GC,z_p,z_c}$
is the grand-canonical ensemble average, and 
 $\left[\cdot \right]$ is the average over the matrix realizations. 
In the two-phase region
the histograms show a double-peak structure. In order to obtain
$z_c^*$ at fixed $z_p$ in a finite volume, several different methods can
be used. We followed two different recipes, the equal-area and the
equal-height methods, as discussed in Ref.~\cite{AP-11}. They give 
completely consistent results: the results we report have been obtained by 
using the equal-height method.

\section{Bulk behavior} \label{sec3}

Before considering the model in the presence of the matrix, we determine the
phase behavior in the bulk. 
We use the algorithm described in Ref.~\cite{AP-11}. One Monte Carlo 
iteration consists in 3 simulated-tempering fugacity swaps and 
1000-5000 grand-canonical moves in which colloids and polymers are inserted or 
removed. For each value of $z_p$, we perform $N_\textrm{ini}$ iterations to 
determine the umbrella functions and then $N_\textrm{iter}$ iterations to 
measure the histograms. Typically, $N_\textrm{ini}$ varies between 
$5000 \, N_m$ and $20000 \, N_m$, while $N_\textrm{iter}$ is of the order
of $5\cdot 10^6 \, N_m$. Here $N_m$ is the number of colloid 
fugacities which are sampled together in the
simulated-tempering simulation; 
we take $N_m\approx 10$. 

\begin{figure} 
\begin{center} 
\includegraphics[width=0.6\textwidth]{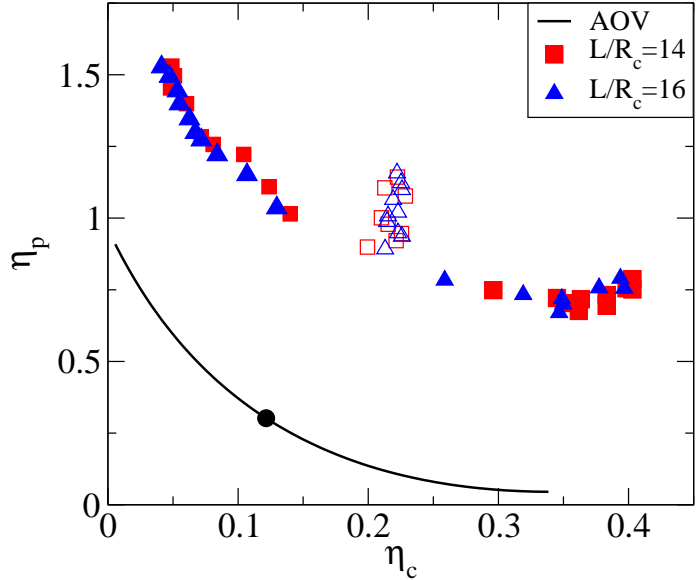}
\caption{(Color online) 
Bulk demixing curve (solid symbols) and coexistence diameter
(empty symbols) in the $(\eta_c,\eta_p)$ system representation 
for $L/R_c = 14, 16$.
We also report the demixing curve of the AOV model, 
solid line, from Ref.~\cite{AP-11}; the solid circle gives the position of the
critical point. \\
}
\label{fig1}
\end{center}
\end{figure}

The demixing curves have been determined for $L/R_c=14$ and $L/R_c=16$,
to identify size effects. We have also performed simulations for $L/R_c = 13$
close to the critical point, to better determine its position.
In Fig.~\ref{fig1} we report the liquid-gas coexistence curve in 
terms of the colloid and polymer volume fractions 
$\eta_c$ and $\eta_p$ defined as
\begin{equation}
 \eta_c = \frac{4}{3} \pi c_c R^3_c , 
  \qquad \eta_p = \frac{4}{3} \pi c_p R^3_g,
\end{equation}
where $c_c$ and $c_p$ are the concentrations of the colloids and of 
the polymers, respectively. It is clear that size effects are quite small,
indicating that our results provide a good estimate of the 
true (infinite-volume) binodal curve. 
As already observed in Ref.~\cite{BLH-02}, the presence of polymer-polymer
interactions significantly changes the binodal curve: 
the values of $\eta_p$ at which demixing occurs
are significantly larger in the presence of polymer interactions 
than in the AOV case. An unusual
feature of our results for the 
coexistence curve is that $\eta_p$ along the coexistence line 
in the colloid-liquid phase slightly increases as $\eta_c$ increases
beyond 0.35. 
This feature is not observed in the more accurate 
model used in Ref.~\cite{BLH-02}, 
nor in experiments. It is probably an artifact of the model, a consequence of 
the discontinuous nature of the potentials or of the fact that polymers
and colloids can overlap paying a relatively small 
energy penalty of the order of 
$2 k_B T$, a phenomenon which is not possible in the exact representation. 

\begin{figure}
\begin{center}
\begin{tabular}{cc}
\includegraphics[width=0.45\textwidth]{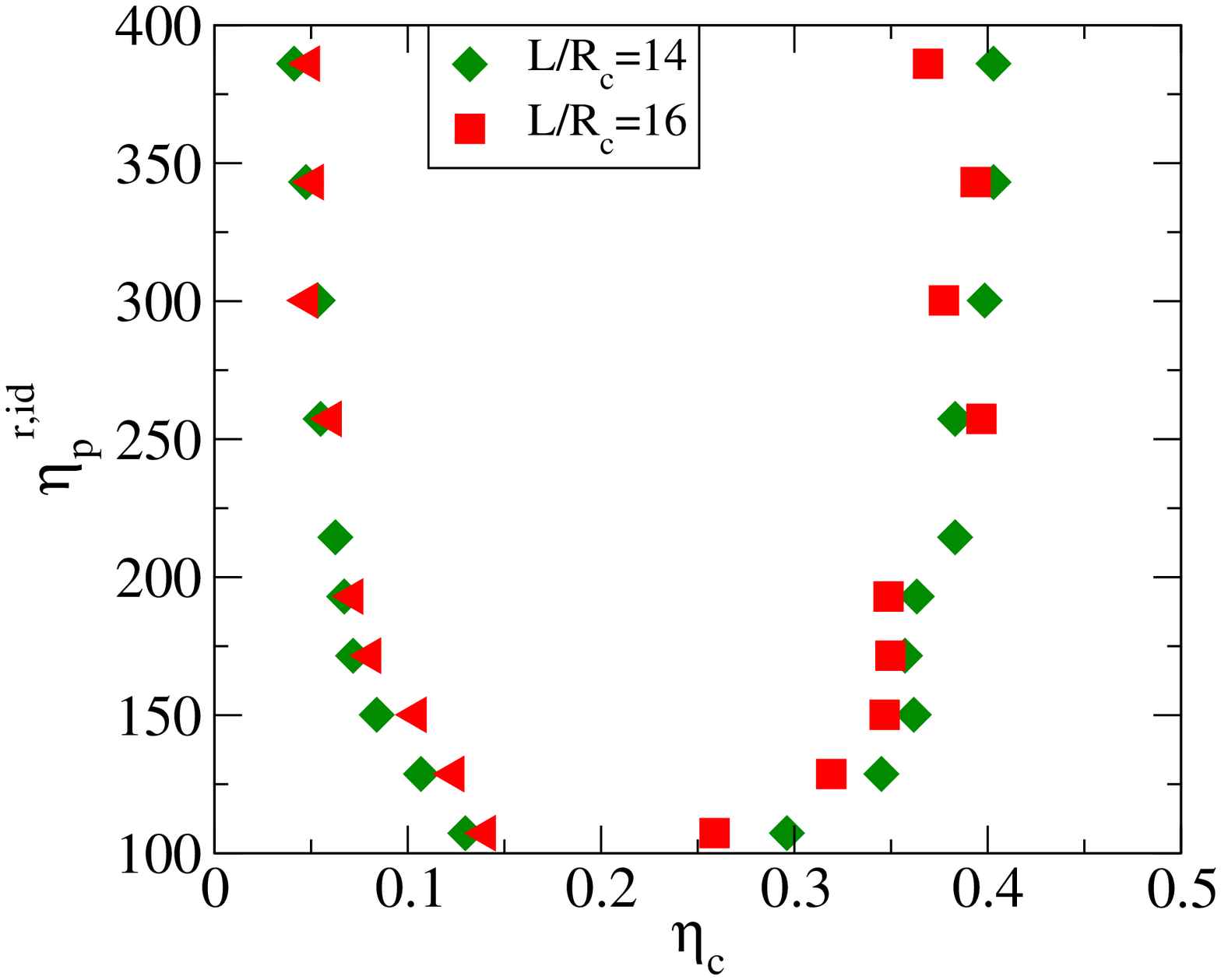} & 
\includegraphics[width=0.45\textwidth]{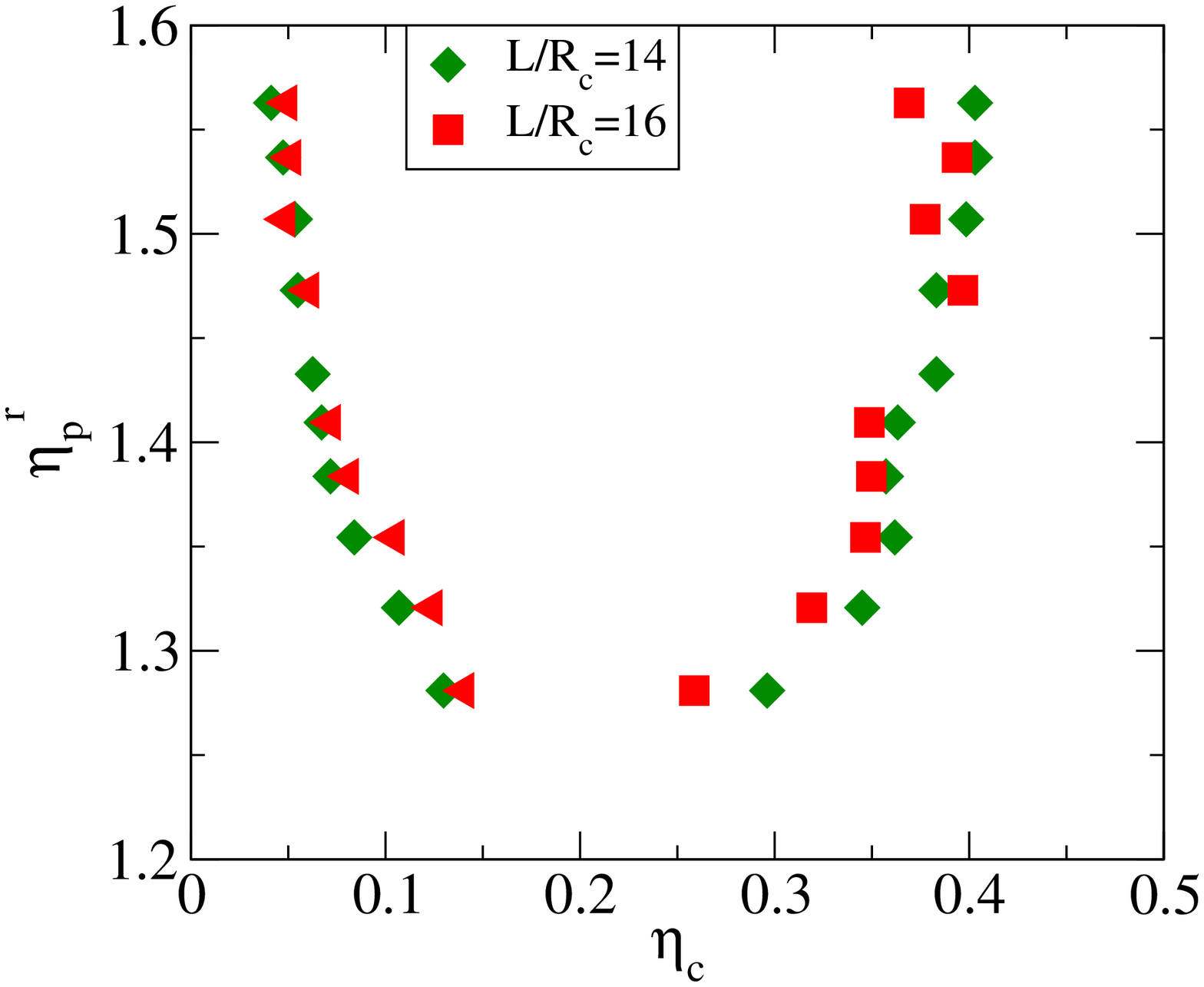} \\
\end{tabular}
\caption{(Color online)
Bulk demixing curve in the reservoir representation for 
$L/R_c = 14, 16$. 
$\eta^{r,{\rm id}}_p$ is the ideal reservoir packing fraction, 
while $\eta^r_p$ is the true reservoir packing fraction.
}
\label{fig:cap5_6}
\end{center}
\end{figure}

We also report the liquid-gas demixing curve in the reservoir representation. 
We report the results in terms of the (ideal) reservoir packing
fraction 
\begin{equation}
 \eta^{r,{\rm id}}_p \equiv \frac{4 \pi}{3} z_p R^3_g,
\end{equation}
which is the quantity used in the AOV model, and in terms of
the reservoir packing 
fraction $\eta^r_p$, which is the volume fraction of the polymers in the 
absence of colloids at a given $z_p$. We have determined it by 
grand-canonical simulations setting $z_c = 0$.
The two quantities $\eta_{p}^{r,\rm id}$ and $\eta_p^r$ are related by
\begin{equation}
\eta_{p}^{r,\rm id} = \eta^r_p \exp[\beta \mu_p^{({\rm exc})}(\eta_p^r)],
\end{equation}
where $\mu_p^{({\rm exc})}(\eta_p^r)$ is the polymer excess  chemical 
potential which can be computed by using the equation of state.
In Fig.~\ref{fig:cap5_6} we report the liquid-gas coexistence curve by using 
both definitions. Size effects are small also in this case, except close to
the critical point. It is interesting to compare the binodal curve in the 
$(\eta_c,\eta_p^r)$ plane with that of the AOV model presented in 
Ref.~\cite{VH-04bis}
(see their Fig.~3). The curve here is significantly more symmetric and shifted 
towards larger values of $\eta_p^r$.

\begin{figure}
\begin{center}
\includegraphics[width=0.6\textwidth]{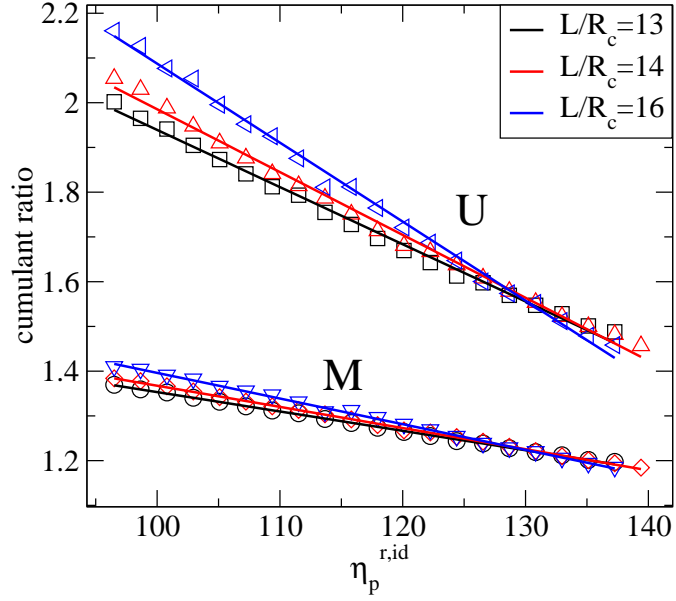}
\caption{(Color online)
Cumulant ratios $M$ and $U$ at coexistence as a function of 
$\eta_p^{r,\rm id}$ for several values of $L/R_c$. }
\label{fig:cap5_7}
\end{center}
\end{figure}

Finally, we determine the critical point by using the standard cumulant 
method (we essentially follow Ref.~\cite{ZVBHV-09}).
The order parameter associated with the critical transition 
can be identified as \cite{ZVBHV-09}
\begin{equation}
 m \equiv \eta_c - \< \eta_c \>,
\end{equation}
where $\< \eta_c \>$ is the average colloid volume fraction. Then, we consider 
the Binder cumulants at coexistence
\begin{equation}
M(z_p,L) \equiv 
\left.\frac{\< m^2 \>}{\< | m | \>^2}\right|_{z_c = z_c^*(z_p)}  \qquad 
U(z_p,L) \equiv 
\left. \frac{\< m^4 \>}{\< m^2 \>^2}\right|_{z_c = z_c^*(z_p)}.
\end{equation}
The cumulants for different values of $L$ are 
reported in Fig.~\ref{fig:cap5_7} as a function of $\eta_p^{r,\rm id}$ for 
three different values of $L/R_c$. 
The critical point is given by the intersection point. Both cumulants
have approximately the same intersection point, which allows us to estimate
\be
 \eta^{r,{\rm id}}_{p,{\rm crit}} = 129(2) , 
  \qquad \eta^r_{p,{\rm crit}} = 1.321(4).
\ee
At the intersection 
$M$ and $U$ assume the values
\be
 M_{\rm crit} = 1.226(5), \qquad U_{\rm crit} = 1.57(2),
\ee
which are close to those appropriate for the Ising three-dimensional
universality class: $M_{\rm crit} \approx 1.239$ \cite{LFP-02} and 
$U_{\rm crit} = 1.6036(1)$ \cite{Hasenbusch-10}. 
The small differences we observe are due to field-mixing effects 
\cite{WB-92,BW-92}, which we have not taken into account in the 
present analysis.
At the critical point we obtain
\begin{eqnarray}
 \eta_{c,{\rm crit}} = \langle \eta_c \rangle = 0.22(1), 
\qquad\qquad
 \eta_{p,{\rm crit}} = \langle \eta_p \rangle = 0.93(2).
\label{crit_bulk}
\end{eqnarray}
Note that these estimates are very different from those of the AOV critical 
point \cite{VH-04bis,VHB-05}:
\begin{equation}
\eta_{c,\rm crit} = 0.1340(2), \qquad\qquad \eta_{p,\rm crit} = 0.3562(6).
\end{equation}
Both the colloid and the polymer critical volume fractions increase 
quite significantly.

The estimate (\ref{crit_bulk}) of the colloid critical volume 
fraction agrees with all experimental and theoretical estimates,
see Table~\ref{tab:cap5_crit} for a list of results. On the other hand, 
the estimate of the polymer volume fraction 
is not consistent with the results of Bolhuis {\em et al.} \cite{BLH-02},
which  used a much more accurate description of the pair interactions.
The discrepancy is probably due to our choice of simplified interactions
and, in particular, to the fact that polymers and colloids can overlap
with a relatively small energy penalty.
Clearly, quantitative predictions require a much more accurate 
modelling of the interactions among polymers and colloids. We also report
some experimental results which show, however, large discrepancies. 
They are of little use for quantitative comparisons. We finally
mention the results of Ref.~\cite{ZVBHV-09}: 
$\eta^{r,\rm id}_{p,\rm crit} = 1.282(2)$, $\eta_{c,\rm crit} = 0.150(2)$,
and $\eta_{p,\rm crit} = 0.328(2)$. As we already discussed,
this model is more appropriate for polymers close to the $\theta$ point
and, indeed, the estimates of the volume fractions at criticality 
are close to the AOV ones. 

\begin{table}
\center
\begin{tabular}{ccccc}
\hline\hline
 {} & {} & $q$ & $\eta_{c,{\rm crit}}$ & $\eta_{p,{\rm crit}}$ \\
\hline
 Expt. & Ref.~\cite{BO-97} & $0.49$ & $0.21(1)$ & $1.00(5)$ \\
 Expt. & Ref.~\cite{IOPP-95}& 0.57 & 0.2 & 0.6 \\
 Expt. & Ref.~\cite{TdK-99} & 0.86  & 0.11      & 0.5 \\
 Expt. & Ref.~\cite{VDDL-96} & $0.92$ & $0.195$ & $1.21$ \\
 Expt. & Ref.~\cite{TSPEALF-08} & 1.00 & 0.2  & 0.5 \\
 MC    & Ref.~\cite{BLH-02} & $0.34$ & $0.20$ & $0.29$ \\
 {} & {} & $0.67$ & $0.19$ & $0.40$ \\
 {} & {} & $1.05$ & $0.18$ & $0.51$ \\
 This work & {} & $0.80$ & $0.22(1)$ & $0.93(2)$ \\
\hline\hline
\end{tabular}
\caption{Experimental (Expt) and Monte Carlo (MC) estimates 
\cite{footnoteCP} of 
$\eta_{c,{\rm crit}}$ and $\eta_{p,{\rm crit}}$. 
An extensive list of results can be found in Sec.~8 of 
Ref.~\protect\cite{FT-08}.
}
\label{tab:cap5_crit}
\end{table}

The critical point can also be approximately determined from the behavior of 
the coexistence diameter given by
\begin{equation}
(\eta_{c,\textrm{diam}},\eta_{p,\textrm{diam}}) \equiv 
\left( \frac{\eta_{c,\textrm{liq}}+\eta_{c,\textrm{gas}}}{2},
\frac{\eta_{p,\textrm{liq}}+\eta_{p,\textrm{gas}}}{2} \right) .
\end{equation}
If we consider the intersection of the diameter with a simple interpolation of 
the coexistence data, we obtain
 $\eta_{c,{\rm crit}} \simeq 0.21$ and $\eta_{p,{\rm crit}} \simeq 0.89$.
The critical colloid fugacity is consistent with 
estimate (\ref{crit_bulk}), while the critical polymer 
fugacity is only slightly underestimated. Size corrections, 
which are not taken into account in this simpler approach, are apparently
small in the bulk, even at the critical point. As we shall see in the next 
section, this is no longer the case in the presence of the matrix.

\section{Demixing in the presence of a porous matrix} \label{sec4}

We now study the demixing in the presence of the matrix for a 
mixture with $q = 0.8$. We take disorder parameters analogous 
to those used in our previous work \cite{AP-11}: we consider
$f = 0.4$ and two different sizes of the quenched colloids, 
$R_{\rm dis} / R_c = 0.2$ and $1.0$. 
Correspondingly, $c_{\rm dis} R^3_c = 0.070, 0.014$ in the two cases,
respectively.
To determine size effects we simulate systems with  $L/R_c = 12, 14, 16$.
We use the same algorithm used in the bulk \cite{AP-11}, setting
$N_\textrm{iter}$, the number of iterations during which histograms 
are measured, equal to $40000 N_m$. Each histogram is averaged over 
400 different matrix realizations. Note that system sizes are smaller than
those used in the AOV case. This is due to the fact that here 
we are dealing with larger polymer volume fractions, hence simulations of the 
interacting model are significantly more time-consuming than AOV simulations
of systems of the same size. This significantly limits the size of the 
systems we can simulate.

\begin{figure}
\begin{center}
\begin{tabular}{cc}
\includegraphics[width=0.45\textwidth]{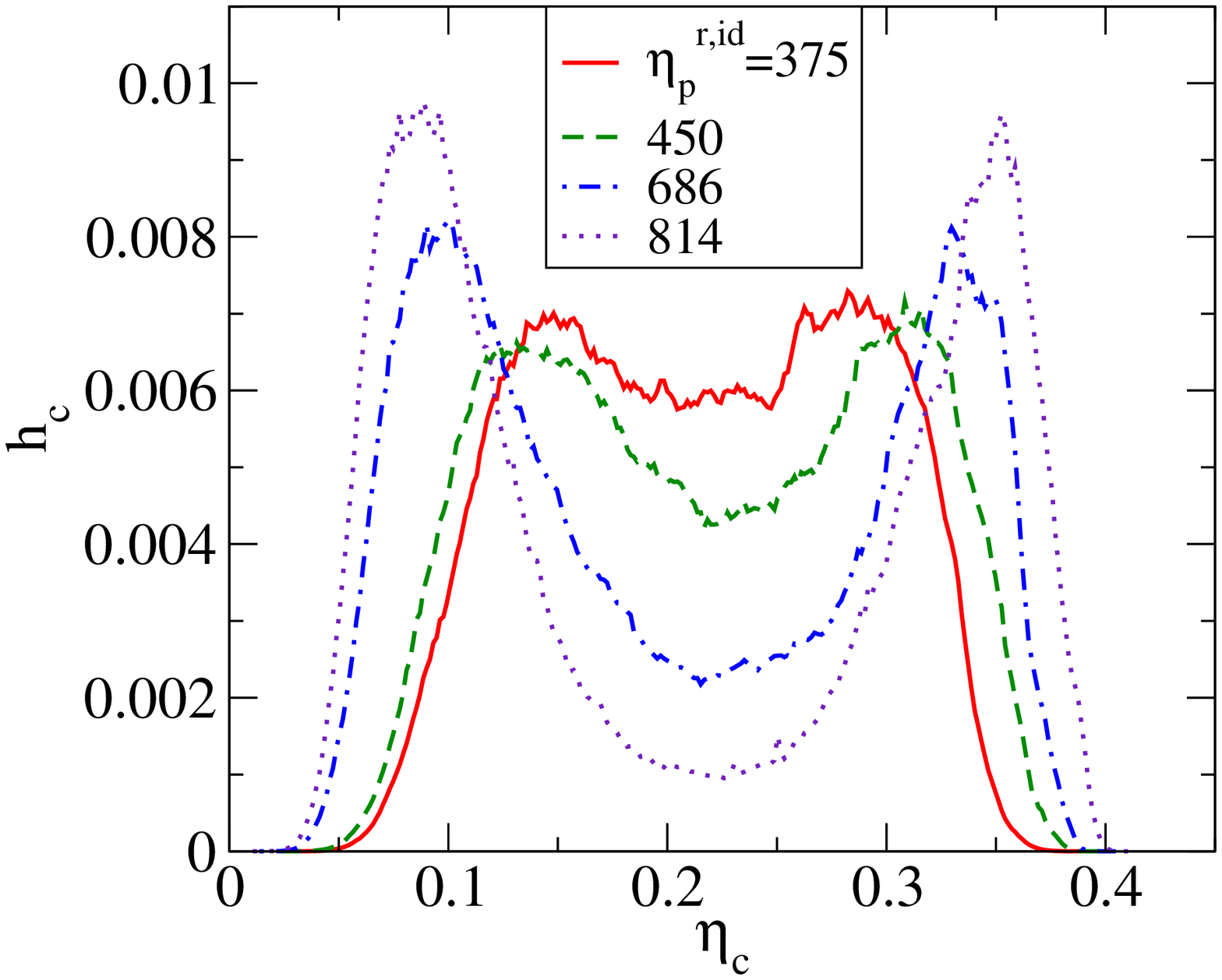} &
\includegraphics[width=0.45\textwidth]{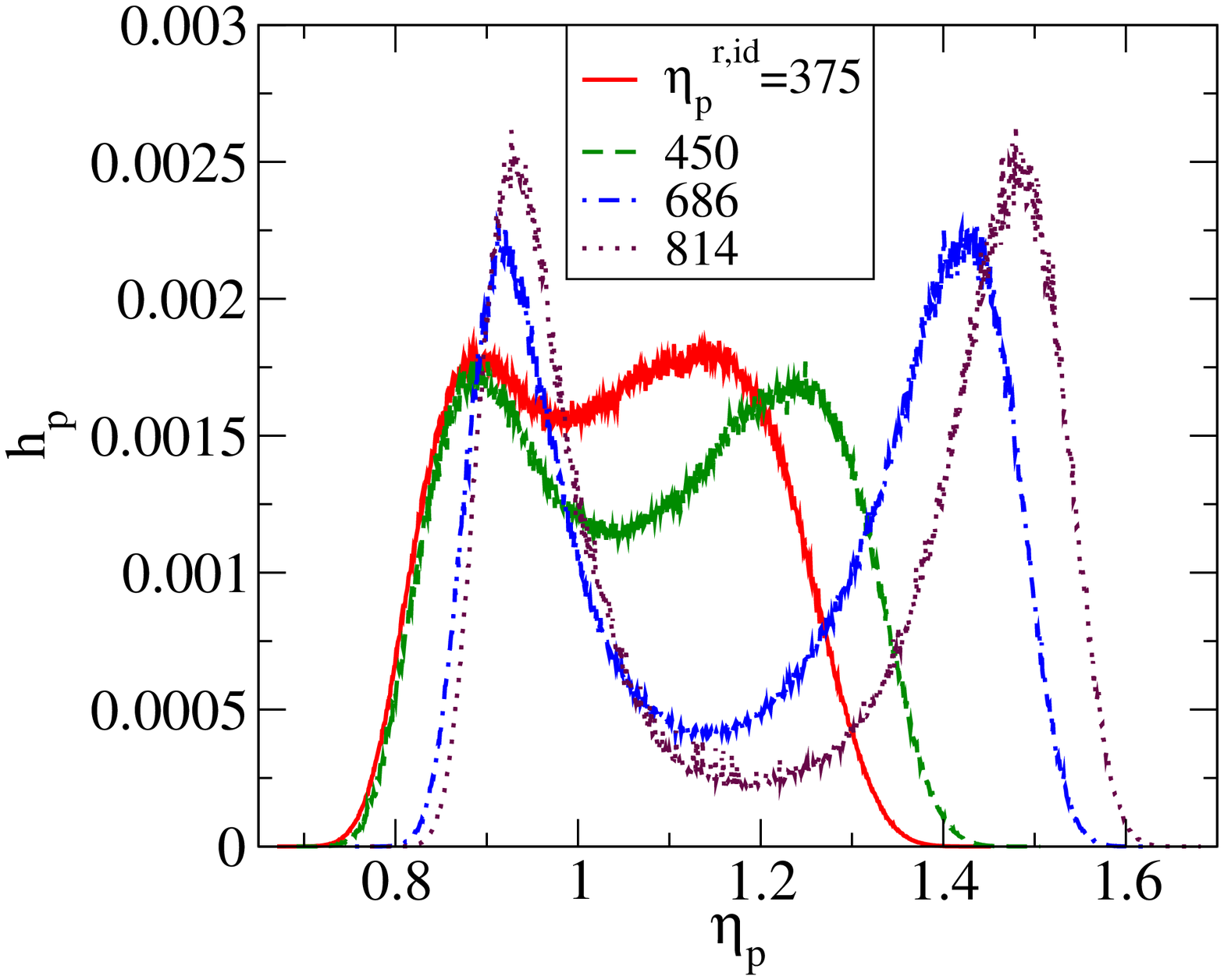} \\
\end{tabular}
\caption{(Color online) 
Colloid (left) and polymer (right) volume fraction histograms at 
coexistence for several values of 
$\eta^{r,\rm id}_p$, for $(f,R_\textrm{dis} / R_c) = 
(0.4,1.0)$, $L/R_c = 14$. 
}
\label{fig:cap6_4}
\end{center}
\end{figure}

\begin{figure}
\center
\begin{tabular}{cc}
\includegraphics[width=0.45\textwidth]{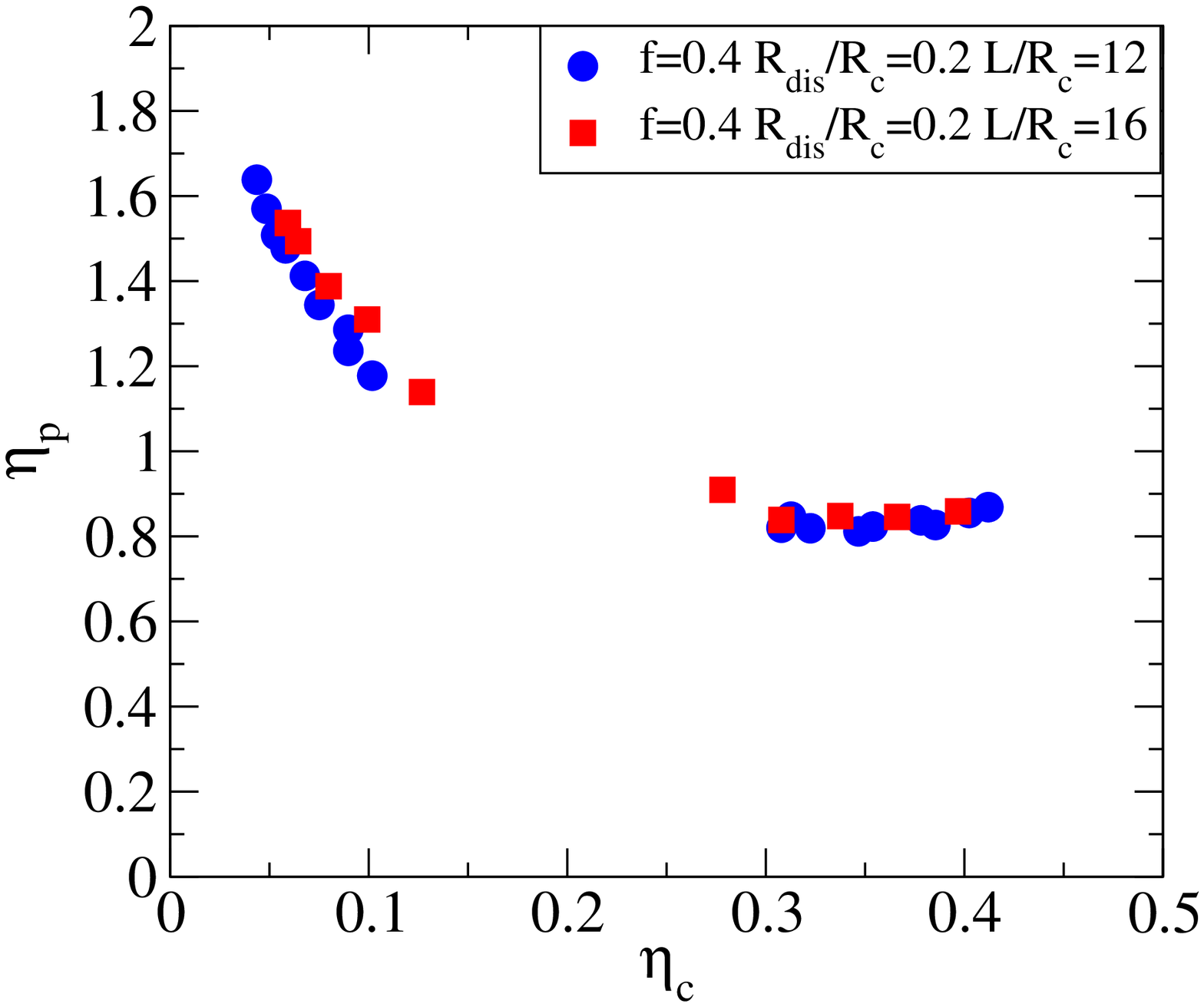} &
\includegraphics[width=0.45\textwidth]{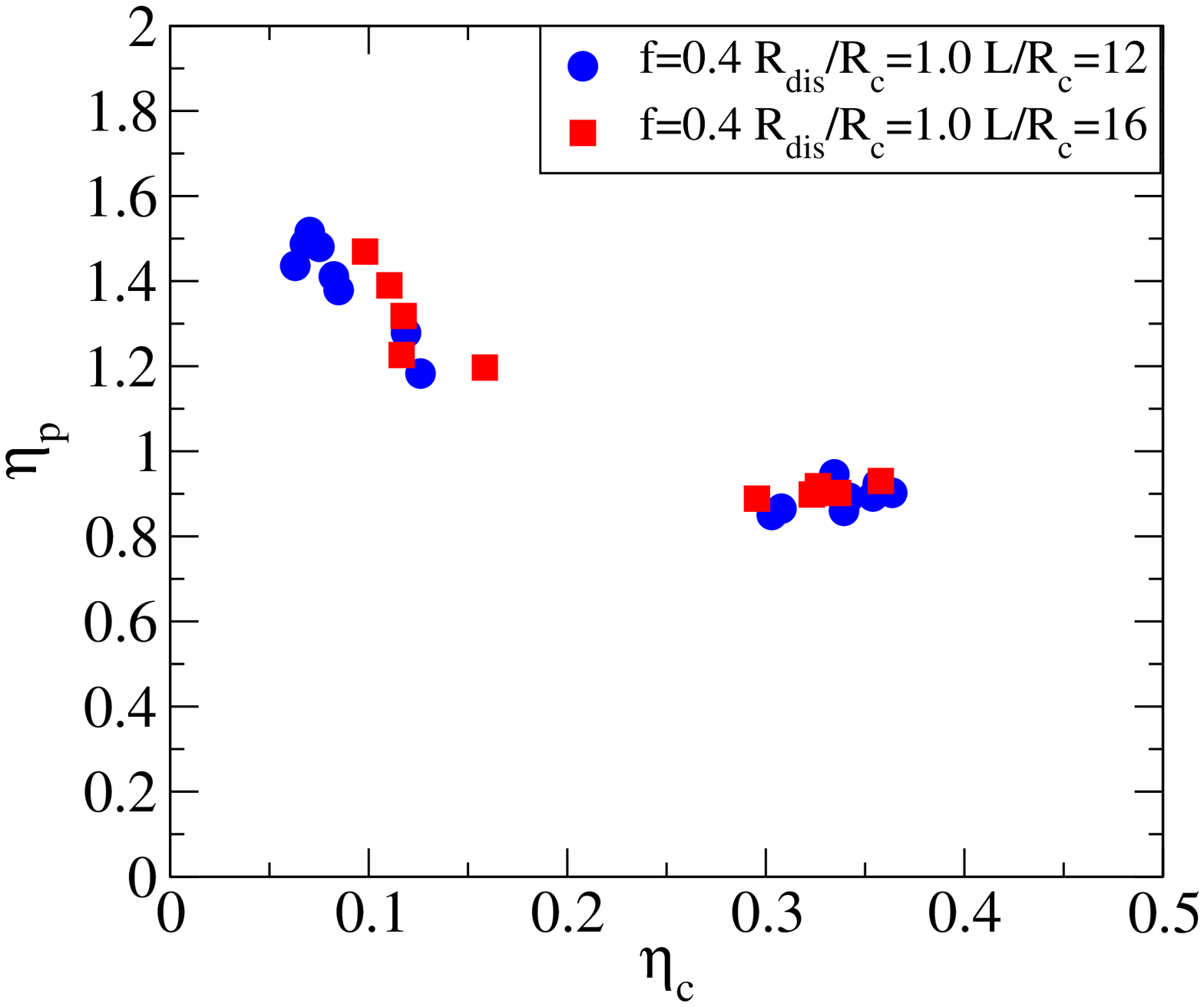} \\
\end{tabular}
\caption{(Color online) 
Fluid-fluid binodal curves for 
$f = 0.4$, $R_{\rm dis}/R_c = 0.2$ (left),
$f = 0.4$, $R_{\rm dis}/R_c = 1.0$ (right). 
We report the results for $L/R_c=12$ and $L/R_c=16$ in terms of 
$\eta_c$ and $\eta_p$.
}
\label{fig:binodal_system_dis}
\end{figure}

\begin{figure}
\begin{tabular}{ll}
\includegraphics[width=0.45\textwidth]{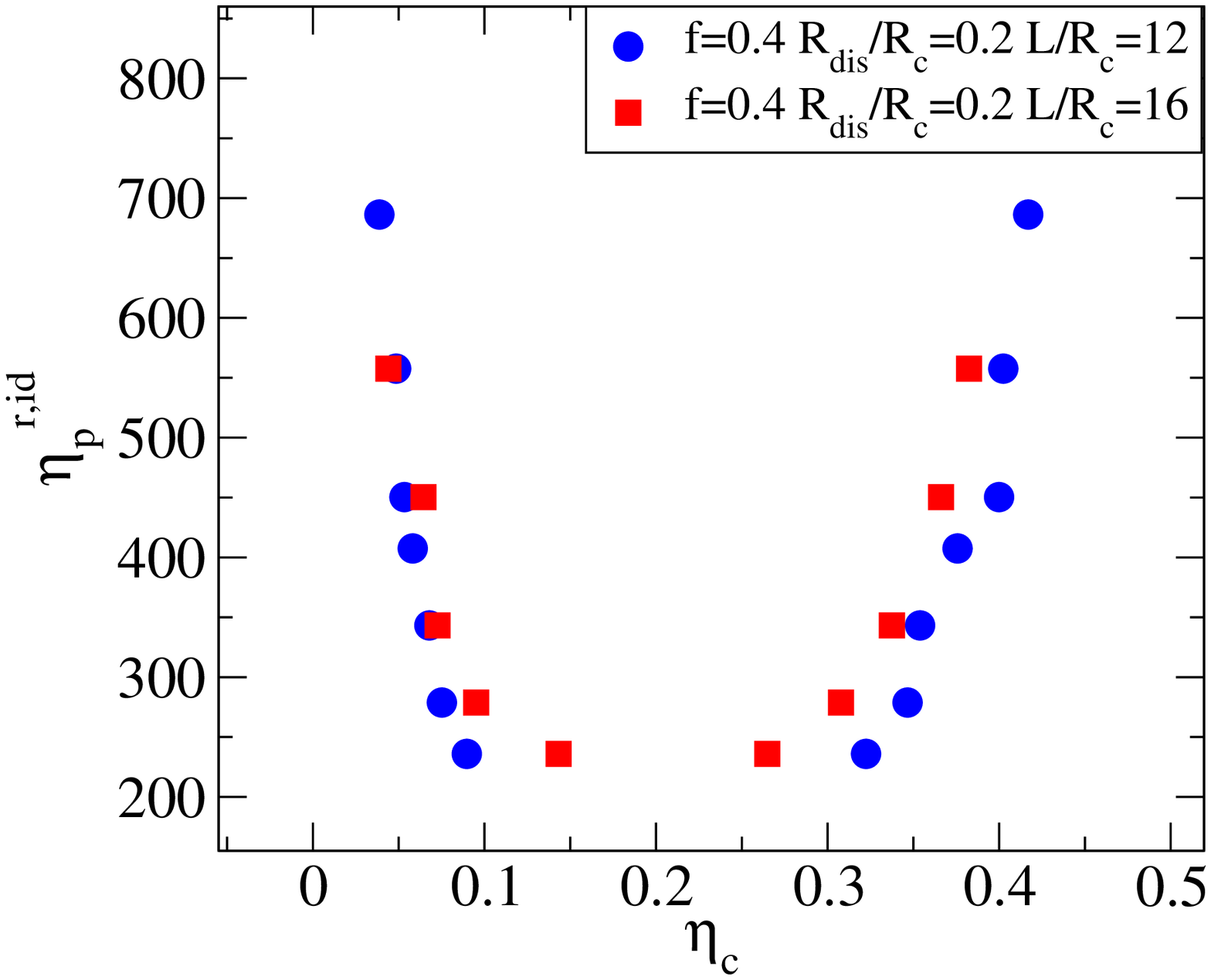}
  \hspace{-0.0truecm} &
\includegraphics[width=0.45\textwidth]{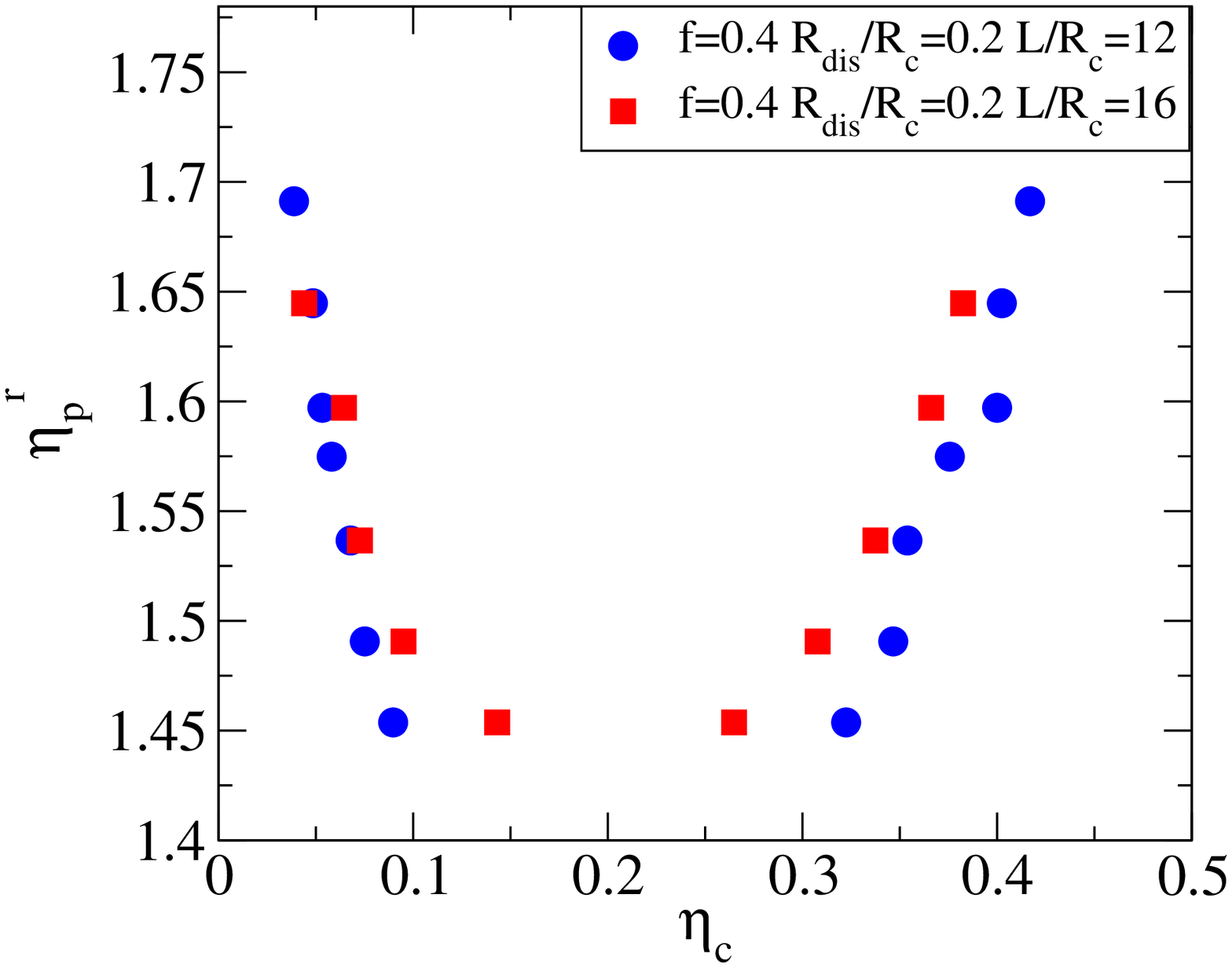} \\
\vspace{0.2cm} \\
\includegraphics[width=0.45\textwidth]{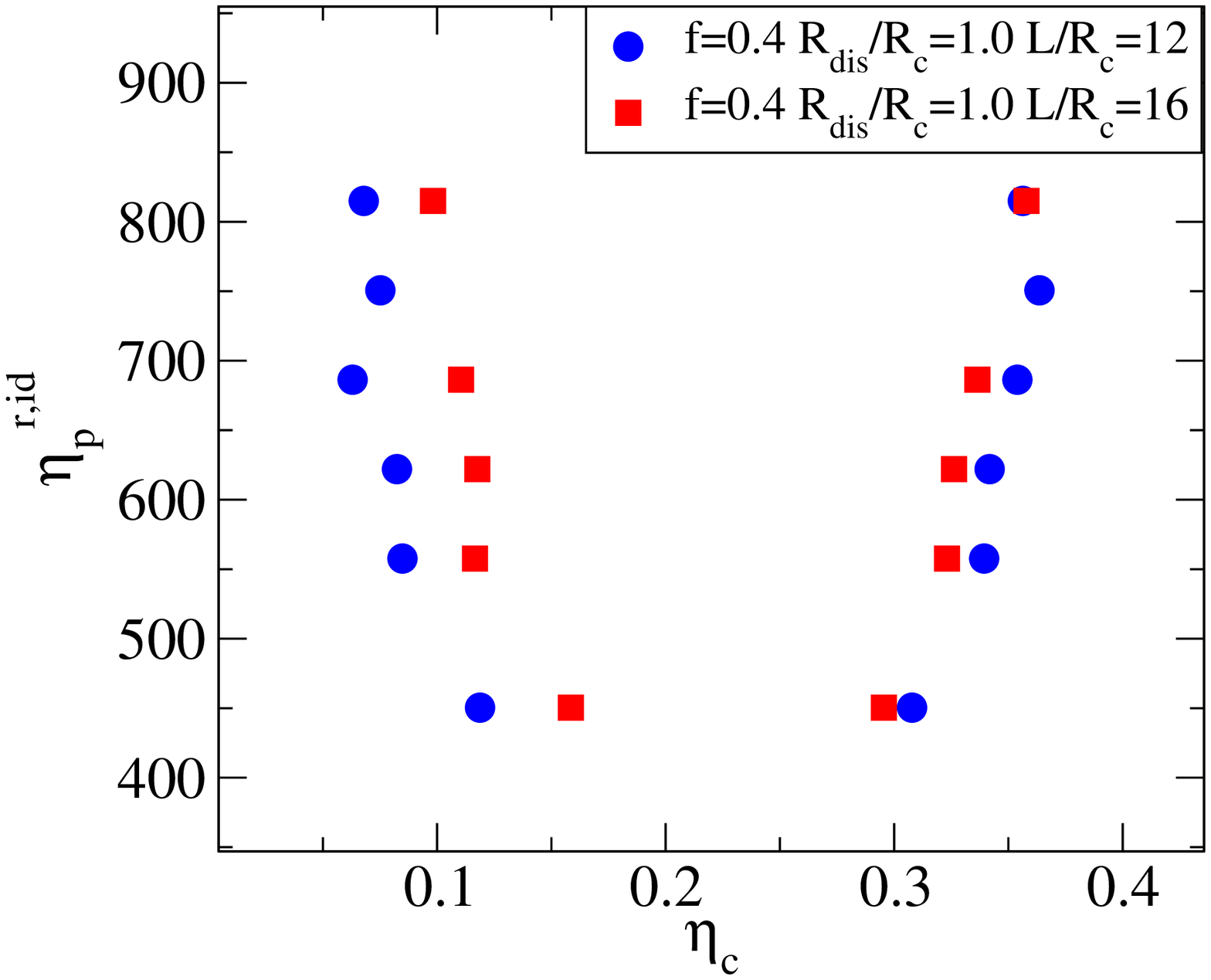}
  \hspace{-0.0truecm} &
\includegraphics[width=0.45\textwidth]{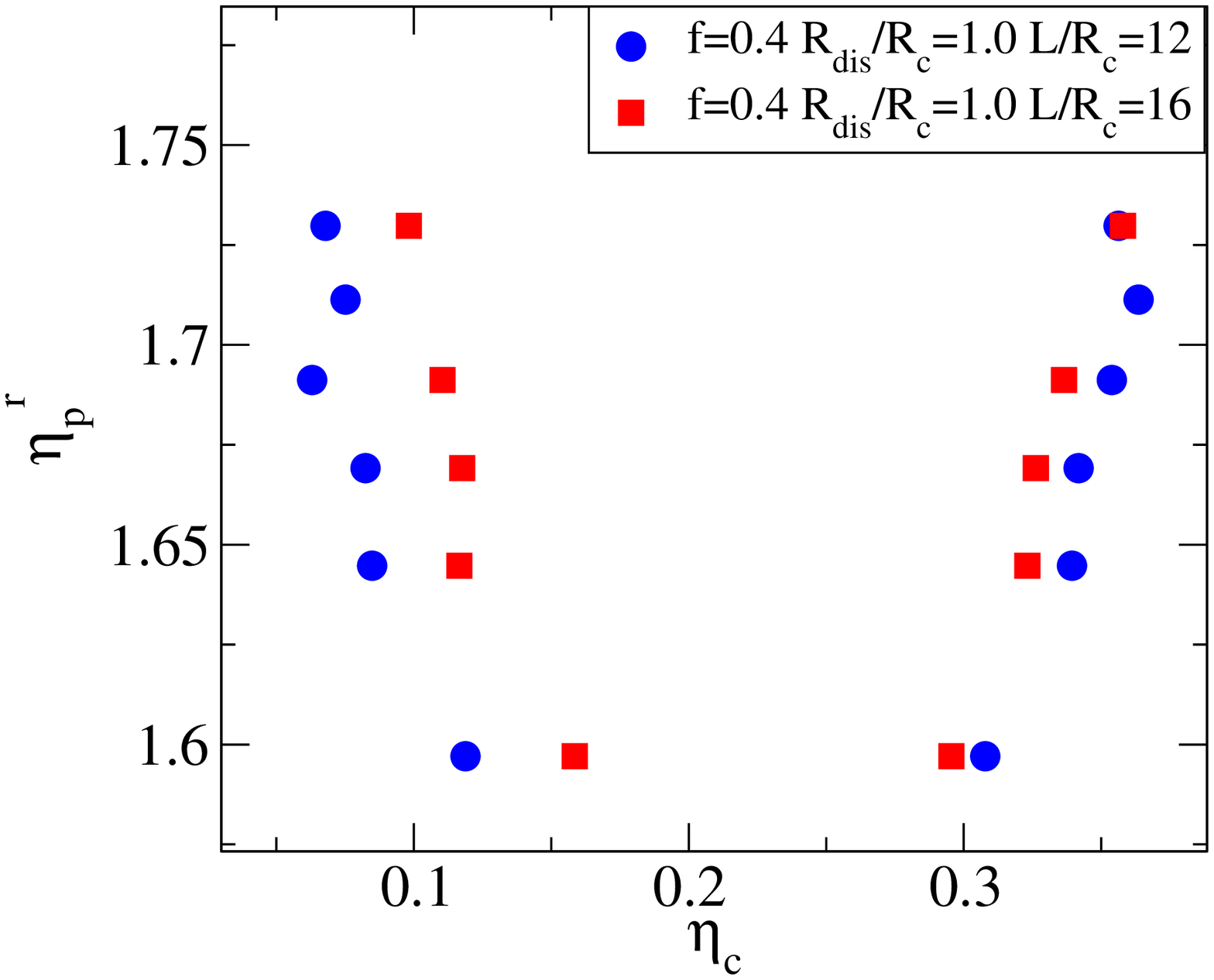} \\
\vspace{0.0cm}
\end{tabular}
\caption{(Color online)
Fluid-fluid binodal curves for $f = 0.4$, $R_{\rm dis}/R_c = 0.2$ (top) and $f = 0.4$, $R_{\rm dis}/R_c = 1.0$ (bottom). 
We report the results for $L/R_c = 12$ and $L/R_c = 16$ in terms of 
$\eta_c, \eta^{r,{\rm id}}_p$ (left) and $\eta_c, \eta^r_p$ (right).
}
\label{fig:binodal_reser_dis}
\end{figure}

\subsection{Demixing curves} \label{sec4.1}

In order to determine the coexistence line $z_c = z_c^*(z_p)$ in the 
$(z_c,z_p)$ plane, we analyze the colloid and polymer averaged histograms
defined in Eq.~(\ref{hphc-def}) using the equal-height method
(completely equivalent results are obtained using the equal-area 
method, see Ref.~\cite{AP-11}): coexistence is defined as the value of 
$z_c$ such that the colloid (or polymer) histogram has two peaks of equal
height. The histograms at coexistence for $R_{\rm dis}/R_c = 1$ are 
shown in Fig.~\ref{fig:cap6_4} for $L/R_c = 14$ 
and several values of $\eta_p^{r,\rm id}$. As expected, 
as $\eta_p^{r,\rm id}$ increases, the two peaks become more 
pronounced, $\eta_{c,\rm gas}$ (the colloid volume fraction in the 
colloid-gas phase) decreases, while $\eta_{c,\rm liq}$ (the same quantity
in the colloid-liquid phase) increases. As in the bulk case, the behavior of the
polymer histograms is more peculiar, since the 
polymer volume fraction apparently increases in both phases, except close 
the the critical point. Again, we expect this to be an artifact of the model,
due to the simplifications we have introduced. 
It is interesting to stress a second difference with respect to the AOV case. 
While the AOV histograms are strongly asymmetric, with a broad colloid-gas
peak and a narrow colloid-liquid peak, in the presence of interactions
the two peaks are much more symmetric. As a consequence, the different 
methods we used to determine coexistence, the equal-height and the 
equal-area methods (see Ref.~\cite{AP-11} for the precise definitions) 
give fully consistent results,
both for the colloid and polymer densities at coexistence and for 
the value $z_c^*$, for all values of $L$.

In Fig.~\ref{fig:binodal_system_dis} we report our results for 
the demixing curves in terms
of $\eta_c$ and $\eta_p$ for two values of $L/R_c$. For 
$R_{\rm dis}/R_c = 0.2$ size corrections are small and thus
our data allow us to determine reliably
the infinite-volume coexistence curve. 
For $R_{\rm dis}/R_c = 1$ we observe instead some size effects on the 
colloid-gas
branch of the binodal: at a given $\eta_c$, the value of $\eta_p$ 
along the binodal increases as $L/R_c$ is increased from 12 to 16. 
On the scale of the figure, the change is small: quantitatively
it amounts to an increase of the order of 5-10\%.
In Fig.~\ref{fig:binodal_reser_dis} we show the fluid binodals in the 
reservoir representation, using both $\eta^{r,\rm id}_p$ and $\eta^r_p$,
as we did in the bulk case. For $R_{\rm dis}/R_c = 0.2$ size 
corrections appear to be under control, except close to the critical 
point. For $R_{\rm dis}/R_c = 1$ corrections are instead larger, 
especially for the colloid-gas branch, which cannot be reliably
determined even 
far from the critical point, where size corrections should be smaller.
Clearly, accurate estimates require much larger values of $L/R_c$. 

\begin{figure}
\begin{center}
\begin{tabular}{cc}
\includegraphics[width=0.45\textwidth]{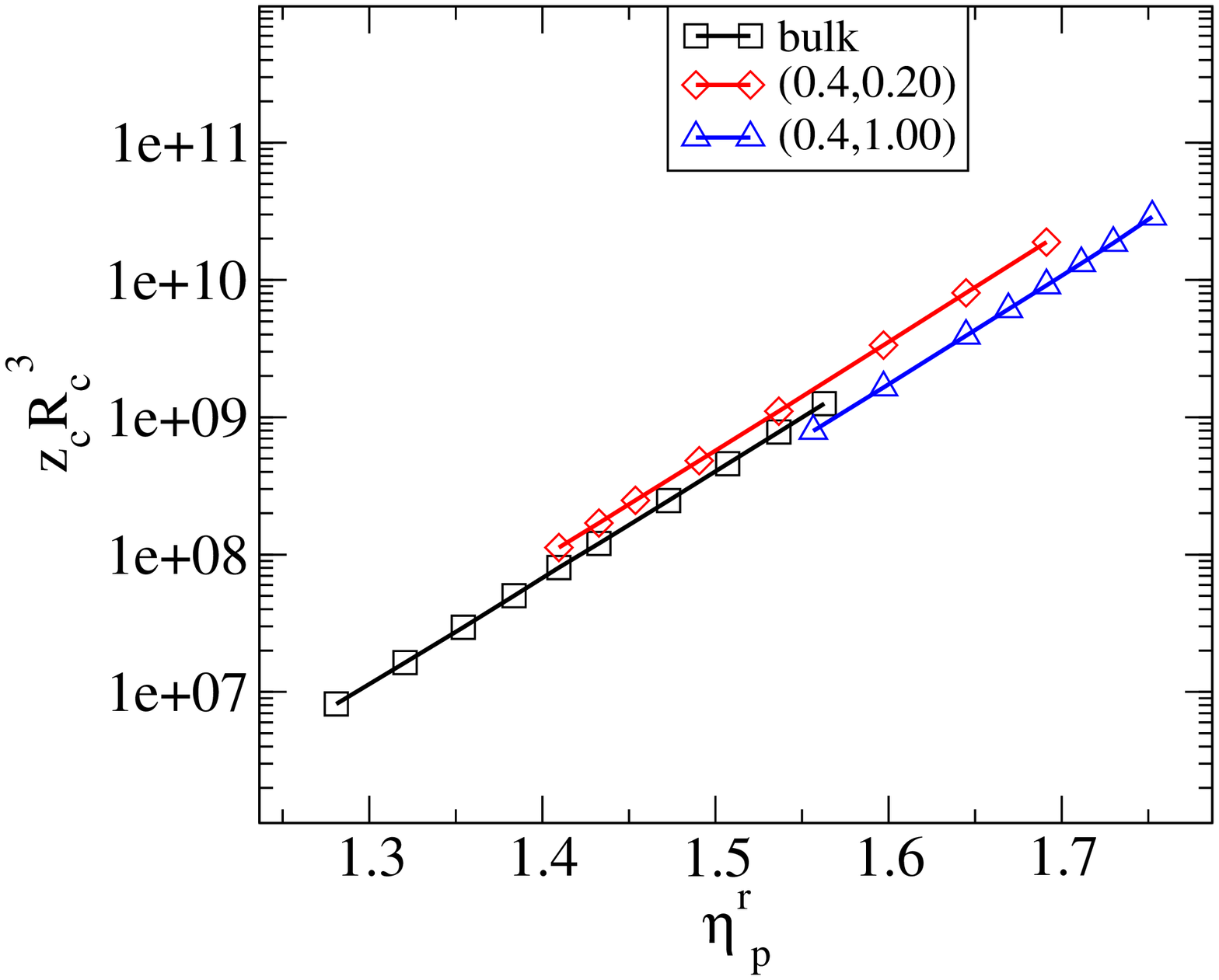} &
\includegraphics[width=0.45\textwidth]{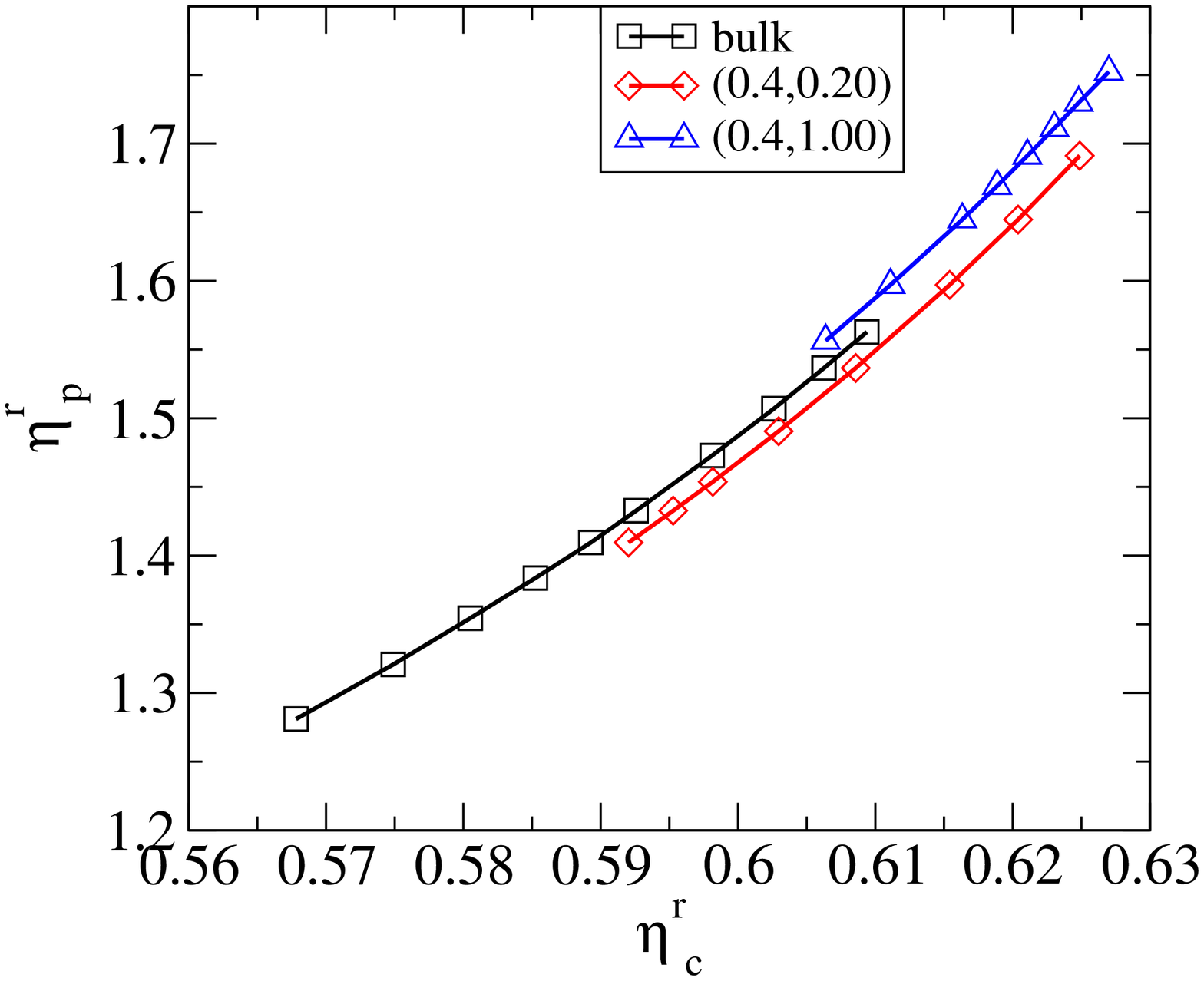} \\
\end{tabular}
\end{center}
\caption{(Color online)
Estimates of $z^*_c R_c^3$ as a function of the reservoir polymer 
volume fraction $\eta_p^r$ (left) and of $\eta_p^{r*}$ at coexistence in terms 
of the reservoir colloid volume fraction $\eta_c^r$ (right). 
Results for $L/R_c = 14$.
}
\label{fig:cap6_z}
\end{figure}

In Fig.~\ref{fig:cap6_z} we report $z^*_c R^3_c$ at coexistence 
as a function of $\eta^r_p$ and also 
$\eta^{r*}_p$ at coexistence in terms of the reservoir colloid
volume fraction $\eta^r_c$, which represents the volume fraction at the 
given value of $z_c$, in the absence of polymers and matrix
\cite{footnote-etarc}.
Note that, on a logarithmic scale, the quantity $z_c^* R_c^3$ 
lies quite precisely on a straight line, indicating that the 
colloid chemical potential at coexistence is well approximated by a linear 
function in $\eta_p^r$. This feature was already observed in the AOV case
and seems to be a general feature of this type of systems.
Fig.~\ref{fig:cap6_z} differs significantly from that obtained in the 
AOV case. There, the binodal curves showed a significant dependence on $f$,
while here the bulk curve approximately falls on top of those 
corresponding to $f=0.4$: they only differ
for the position of the critical point. This implies that a 
colloid-liquid bulk phase is almost always in equilibrium with a colloid-liquid phase
in the matrix, and so does a colloid-gas phase: no capillary condensation or 
evaporation is observed except in a very tiny parameter range, i.e. 
for those ($\eta_c^r$, $\eta_p^r$) that belong to the tiny region between
the bulk and matrix binodal curve.

\subsection{Critical point} \label{sec4.2}

We wish now to estimate the position of the critical points. 
This is not an easy task in these
systems, since the transition belongs to the universality class of the 
random-field Ising model (RFIM) 
\cite{VBL-06,VBL-08,PVCL-08,Vink-09}. Size corrections are large and 
the efficient cumulant method we used for the bulk case does not work. 
Therefore, we must adopt a different strategy: we 
follow closely Ref.~\cite{VBL-08}.
First, for each matrix realization we define the averages 
$\langle N_c^k\rangle_{\rm vap}$ and $\langle N_c^k\rangle_{\rm liq}$ 
in the vapor and liquid phases. They are defined as 
\begin{eqnarray}
\langle N_c^k\rangle_{\rm vap} &= &
   {1\over K_{\rm vap}} \int \theta(N_{c,\rm int} - N_c) P(N_c) N_c^k ,
\nonumber \\
\langle N_c^k\rangle_{\rm liq} &= &
   {1\over K_{\rm liq}} \int \theta(N_c - N_{c,\rm int}) P(N_c) N_c^k .
\end{eqnarray}
Here $P(N_c)$  is the histogram of $N_c$ at coexistence for the given 
matrix realization, $N_{c,\rm int}$ gives the position of the minimum 
between the two peaks, $\theta(x)$ is Heaviside step function
[$\theta(x) = 1$ for $x > 0$ and $\theta(x) = 0$ for 
$x < 0$], and $K_{\rm vap}$ and $K_{\rm liq}$ are normalization factors. 
Then, for each phase, we define the connected 
susceptibility \cite{VBL-08}
\begin{equation}
 \chi  = \frac{\left[ \langle N_c^2 \rangle - \langle N_c \rangle^2 \right]}{V},
\end{equation}
where $[\cdot]$ is the average over disorder and $V$ is the volume of the box. 
Essentially, $\chi_{\rm vap}$ and $\chi_{\rm liq}$ measure the 
average squared width of the liquid and vapor peaks for each given sample.

\begin{figure}
\begin{tabular}{ll}
\includegraphics[width=0.45\textwidth]{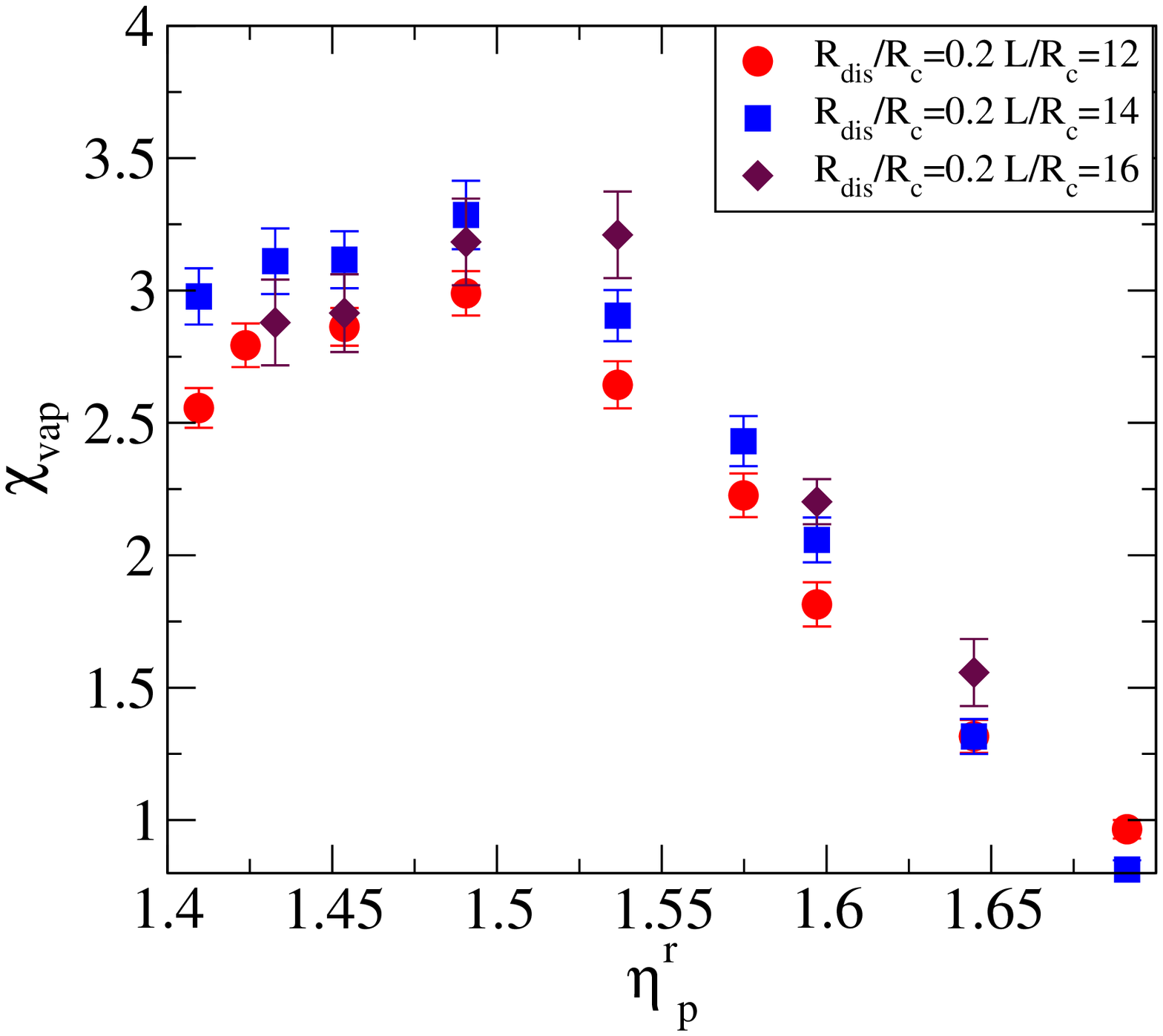}
  \hspace{-0.0truecm} &
\includegraphics[width=0.45\textwidth]{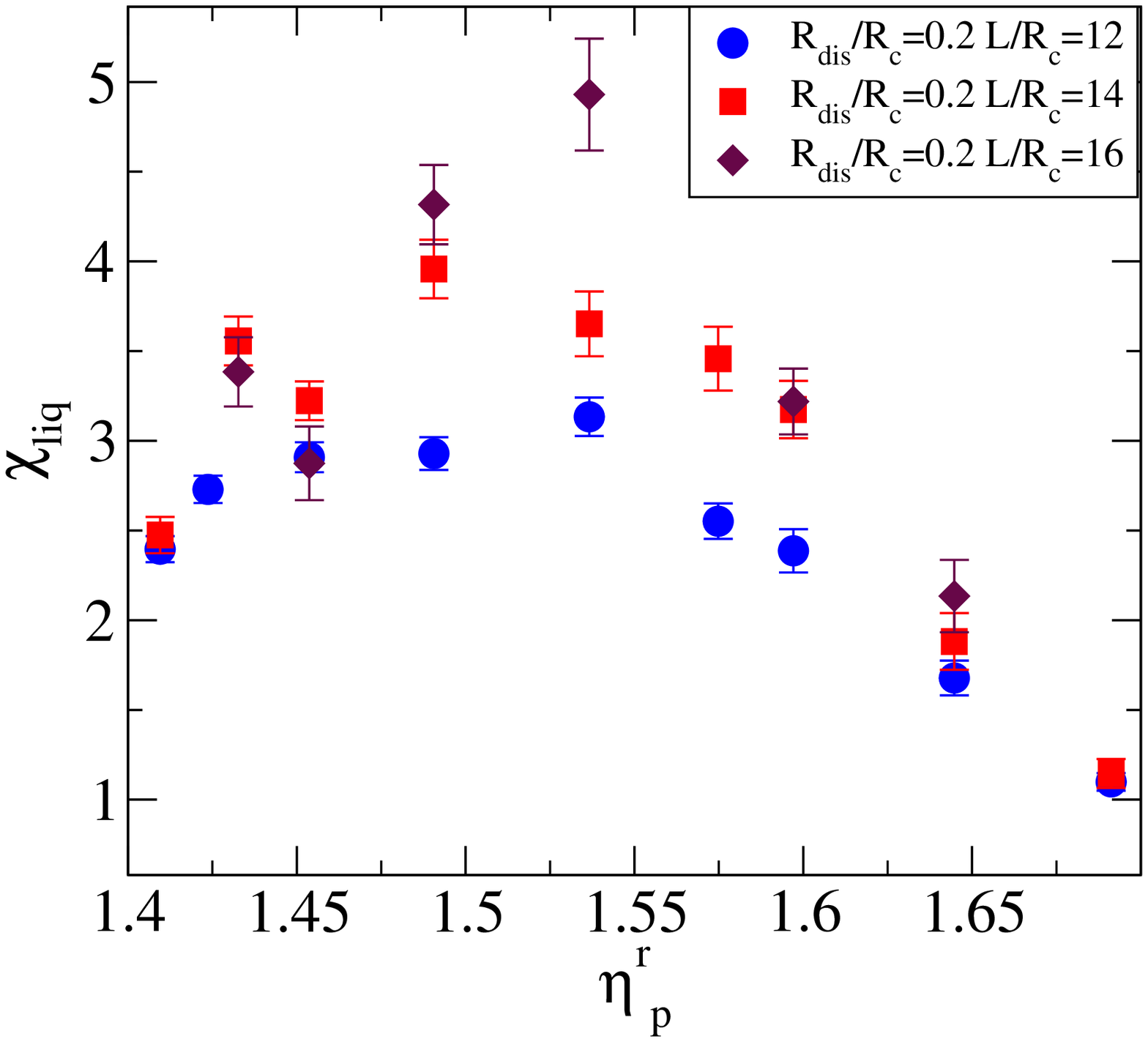} \\
\vspace{0.2cm} \\
\includegraphics[width=0.45\textwidth]{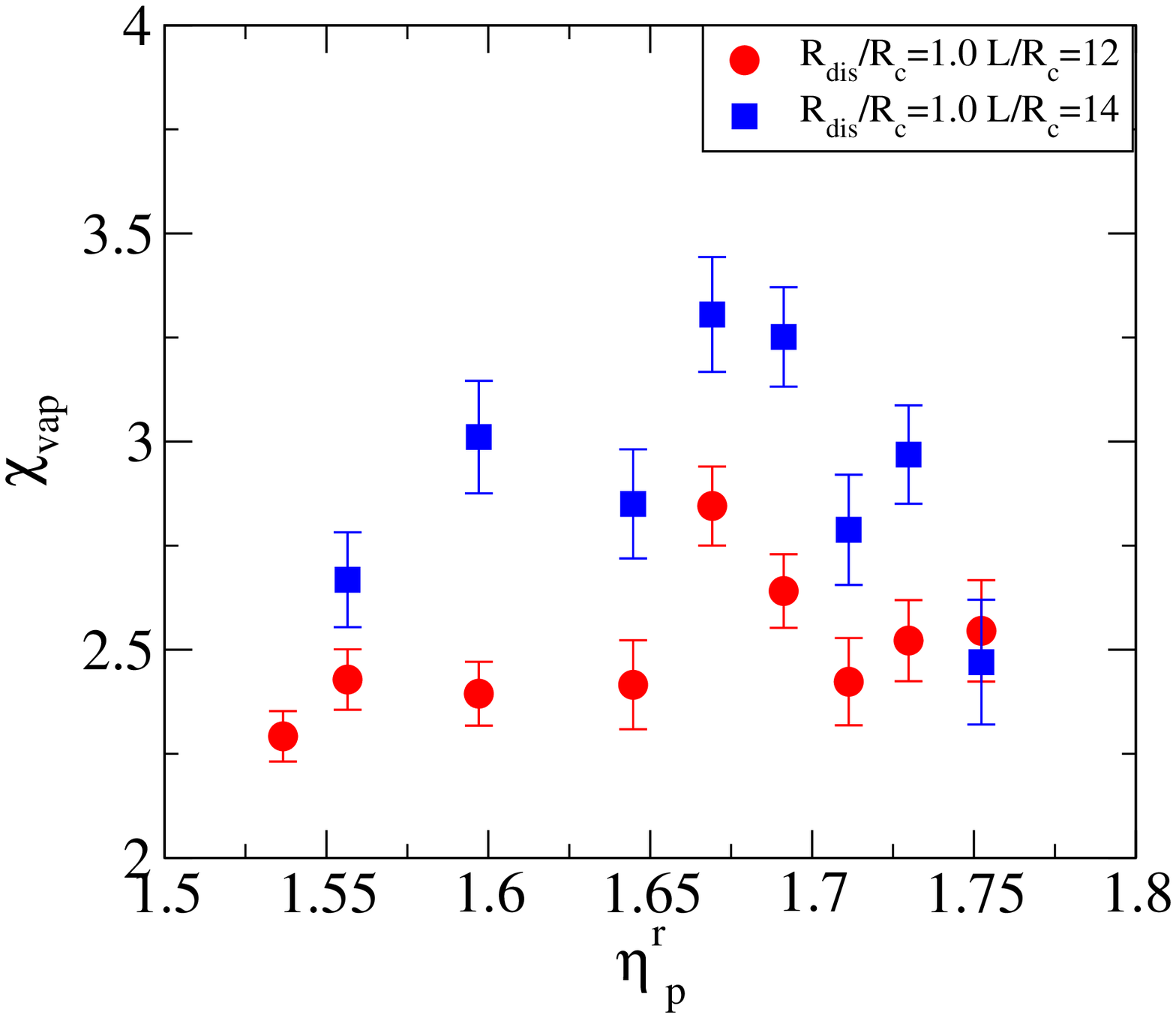}
  \hspace{-0.0truecm} &
\includegraphics[width=0.45\textwidth]{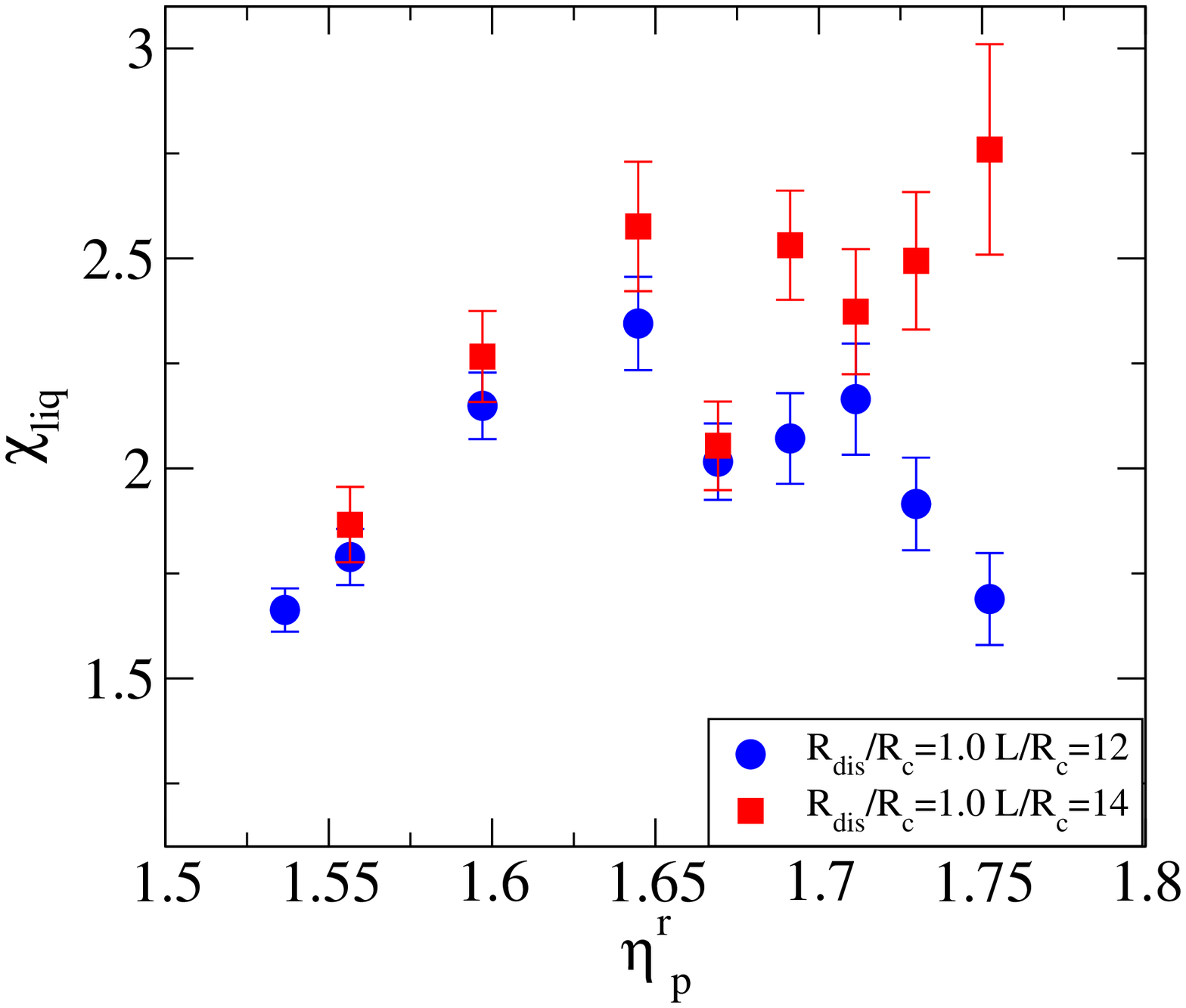} \\
\vspace{0.0cm}
\end{tabular}
\caption{(Color online) 
Connected susceptibilities $\chi$ in
the colloid-vapor (left) and in the colloid-liquid (right) phases for 
$f = 0.4$, $R_{\rm dis}/R_c = 0.2$ (top) and $R_{\rm dis}/R_c = 1$ (bottom). 
We report the results for $L/R_c = 12, 14, 16$ as a function of $\eta^r_p$.}
\label{fig:cap6_chi}
\end{figure}

In Fig.~\ref{fig:cap6_chi} we report the connected susceptibilities 
for $f = 0.4$, $R_{\rm dis}/R_c = 0.2, 1$ and $L/R_c = 12, 14, 16$. 
While the results for $R_{\rm dis}/R_c = 0.2$ are reasonably 
smooth, the data for $R_{\rm dis}/R_c = 1$ are scattered with quite large
error bars (we do not even report the data at $L/R_c = 16$,
since they would make the figure unreadable). 
This is probably due to the small number of samples 
we are using. Indeed, in Ref.~\cite{VBL-08} it was shown that 
sample-to-sample fluctuations may be quite large and require a particularly
large number of different matrix realizations to be controlled. In their
case, they averaged over 2000-3000 different realizations, a number 
which is significantly larger than ours: we only have 400 different samples.
In spite of the large errors,
the data are, at least qualitatively, in agreement with the 
expected behavior: $\chi(L,\eta_p^r)$ increases with $L$ at 
fixed $\eta_p^r$, while, at fixed $L$, it first increases and then decreases
as a function of $\eta_p^r$, as also observed in Ref.~\cite{VBL-08}. 

General renormalization-group arguments indicate that, close to the 
critical point, the susceptibility satisfies the 
finite-size scaling Ansatz
\begin{equation}
\chi = (L/R_c)^{\gamma/\nu} F[t (L/R_c)^{1/\nu}],
\label{FSS}
\end{equation}
where $\gamma$ and $\nu$ are critical exponents, $F(x)$ is a scaling 
function, and $t \equiv \eta^r_p / \eta^r_{p,{\rm crit}} - 1$
measures the ``distance" from the critical point. This 
scaling Ansatz is, strictly speaking, valid in the limit 
$L\to \infty$, $t\to0$ at fixed $t (L/R_c)^{1/\nu}$ and neglects
(analytic and nonanalytic) scaling corrections.

\begin{table}[t]
\begin{center}
\begin{tabular}{cccc}
\hline
\hline
{} & $\nu$ & $\eta$\\
\hline
Newman \& Barkema \cite{NB-96} & 1.02(6)   & 0.15(7) \\
Hartmann \& Young \cite{HY-01} & $1.32(7)$ & $0.50(3)$ \\
Middleton \& Fisher \cite{MF-02} & $1.37(9)$ & $0.51(3)$ \\
Vink {\em et al.} \cite{VBL-08} & $1.1$ & $0.13$ \\
Fernandez {\em et al.} \cite{FMY-11} & $0.90(15)$ & $0.531(40)$ \\
\hline
\hline
\end{tabular}
\end{center}
\caption{Estimates of the critical exponents 
$\nu$ and $\eta$ for the RFIM universality class.
The exponent $\gamma$ is related to $\nu$ and $\eta$ by $\gamma/\nu=2-\eta$.}
\label{tab:RFIM}
\end{table}

\begin{figure}
\begin{tabular}{ll}
\includegraphics[width=0.45\textwidth]{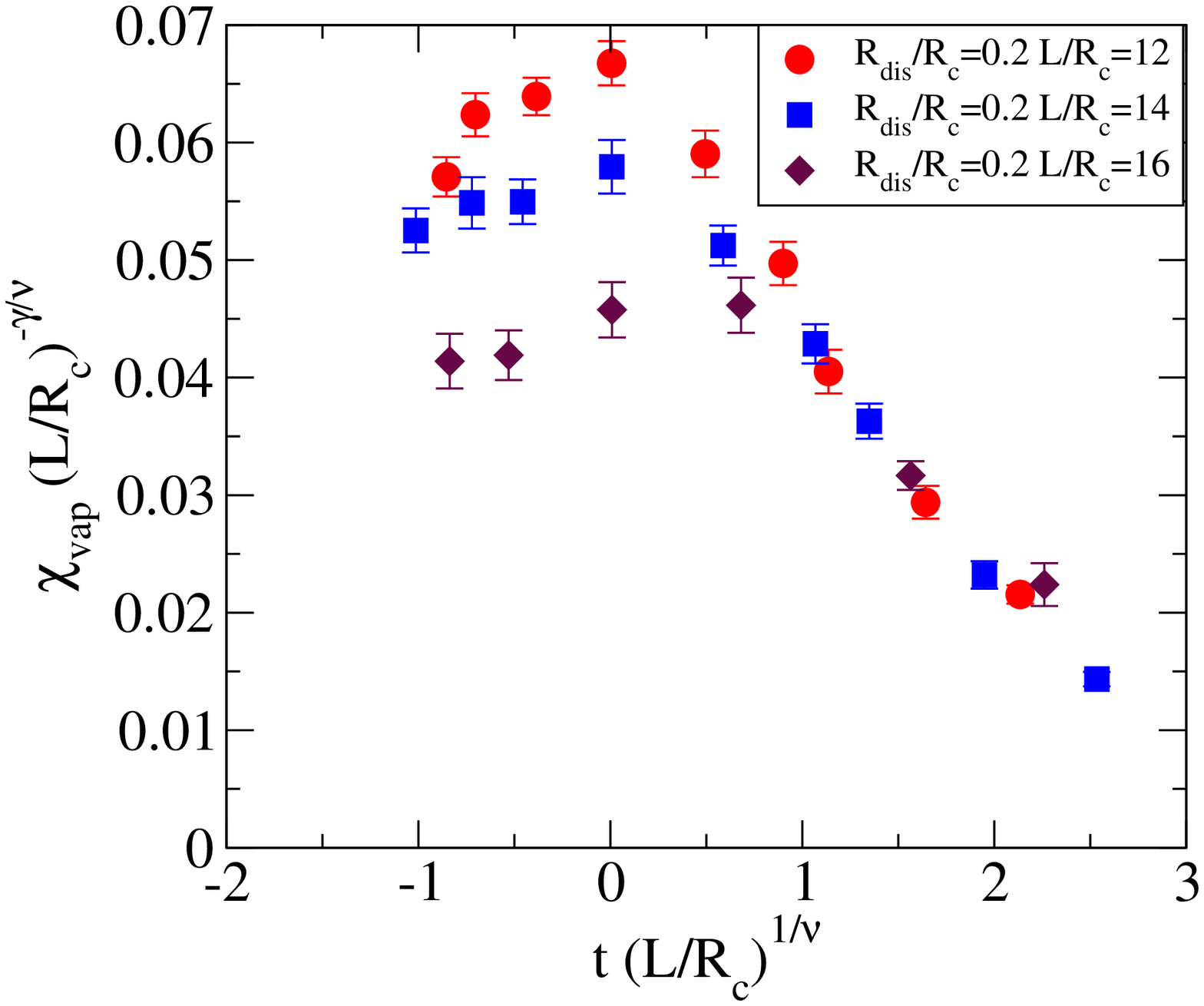}
  \hspace{-0.0truecm} &
\includegraphics[width=0.45\textwidth]{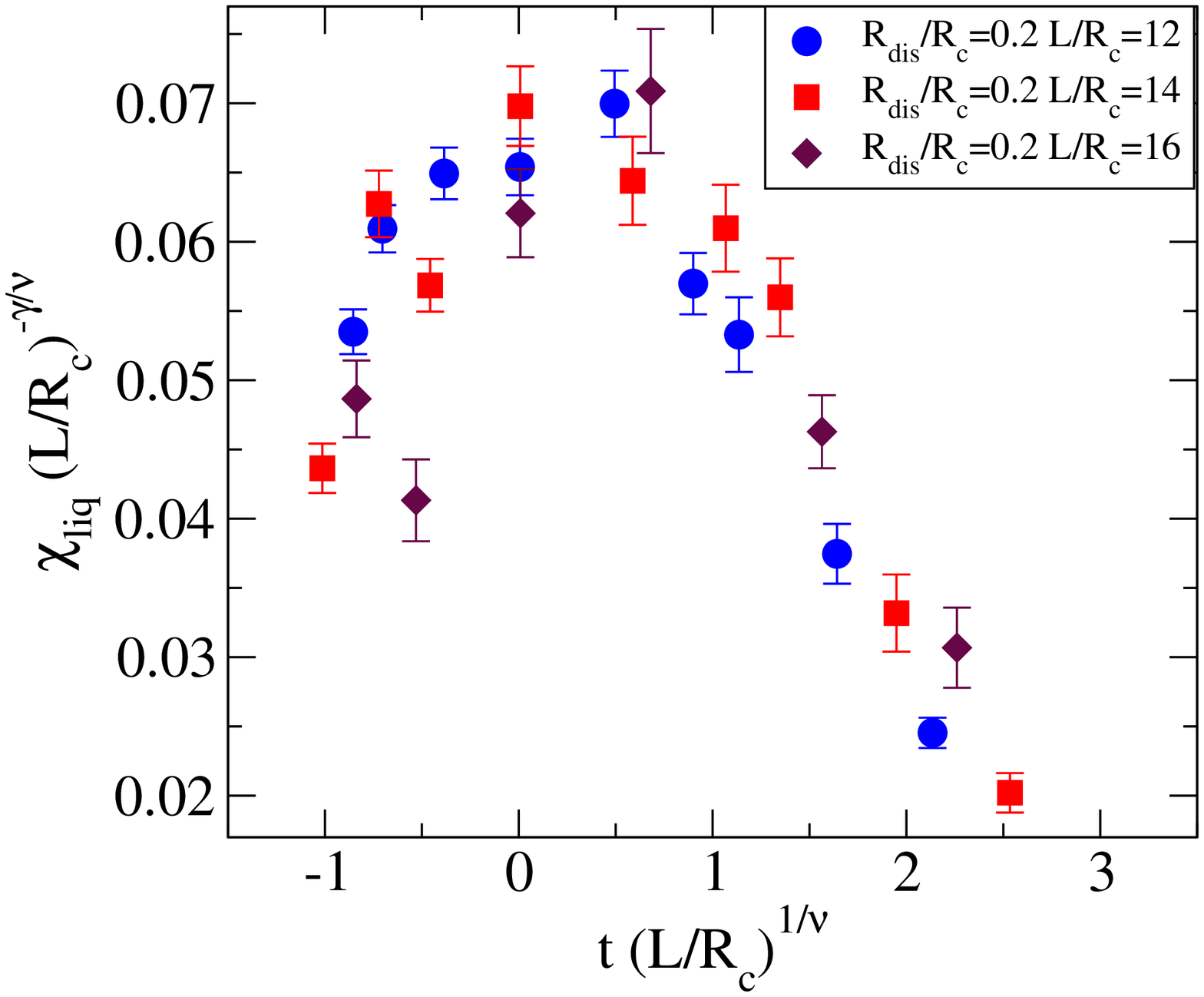} \\
\vspace{0.2cm} \\
\includegraphics[width=0.45\textwidth]{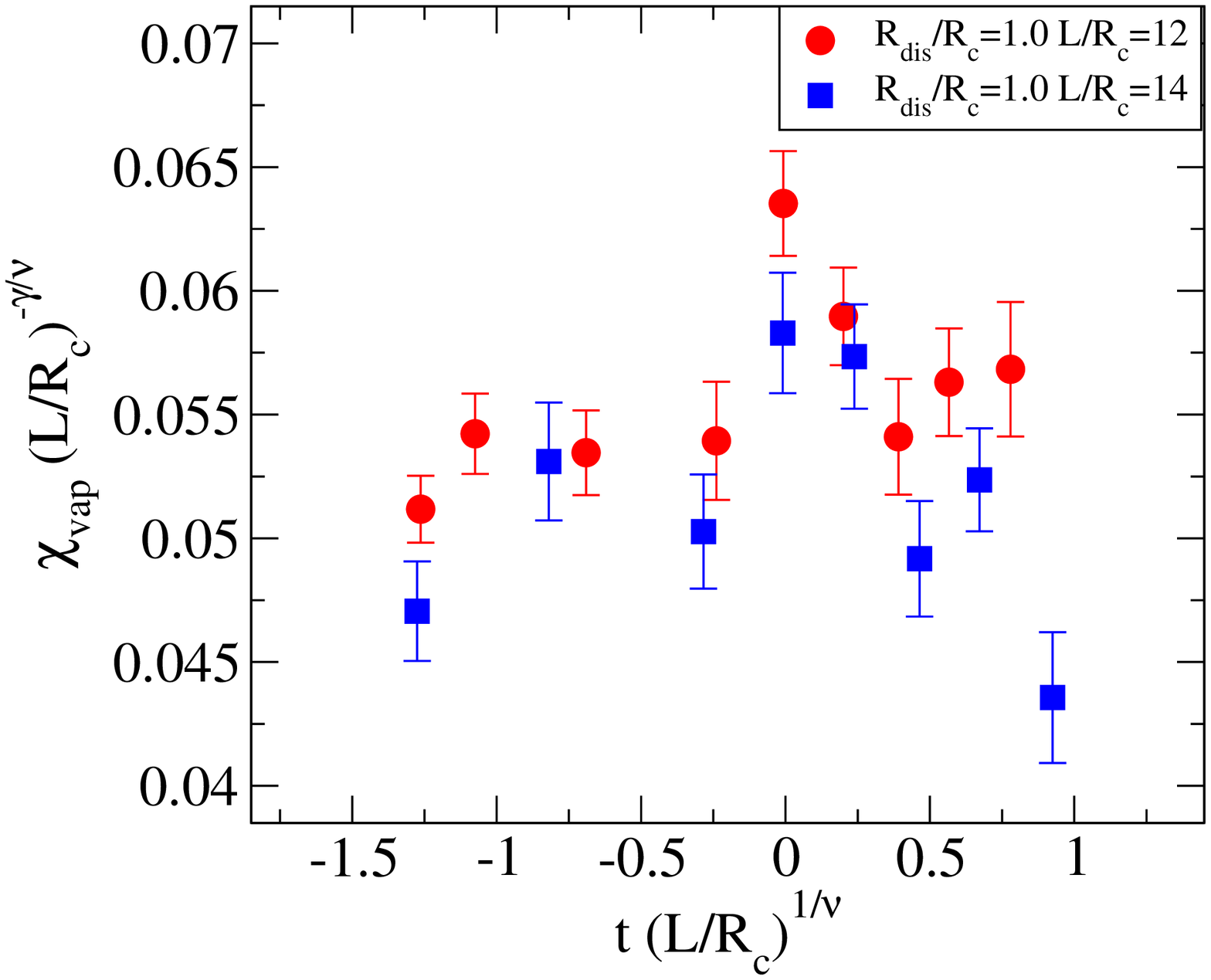}
  \hspace{-0.0truecm} &
\includegraphics[width=0.45\textwidth]{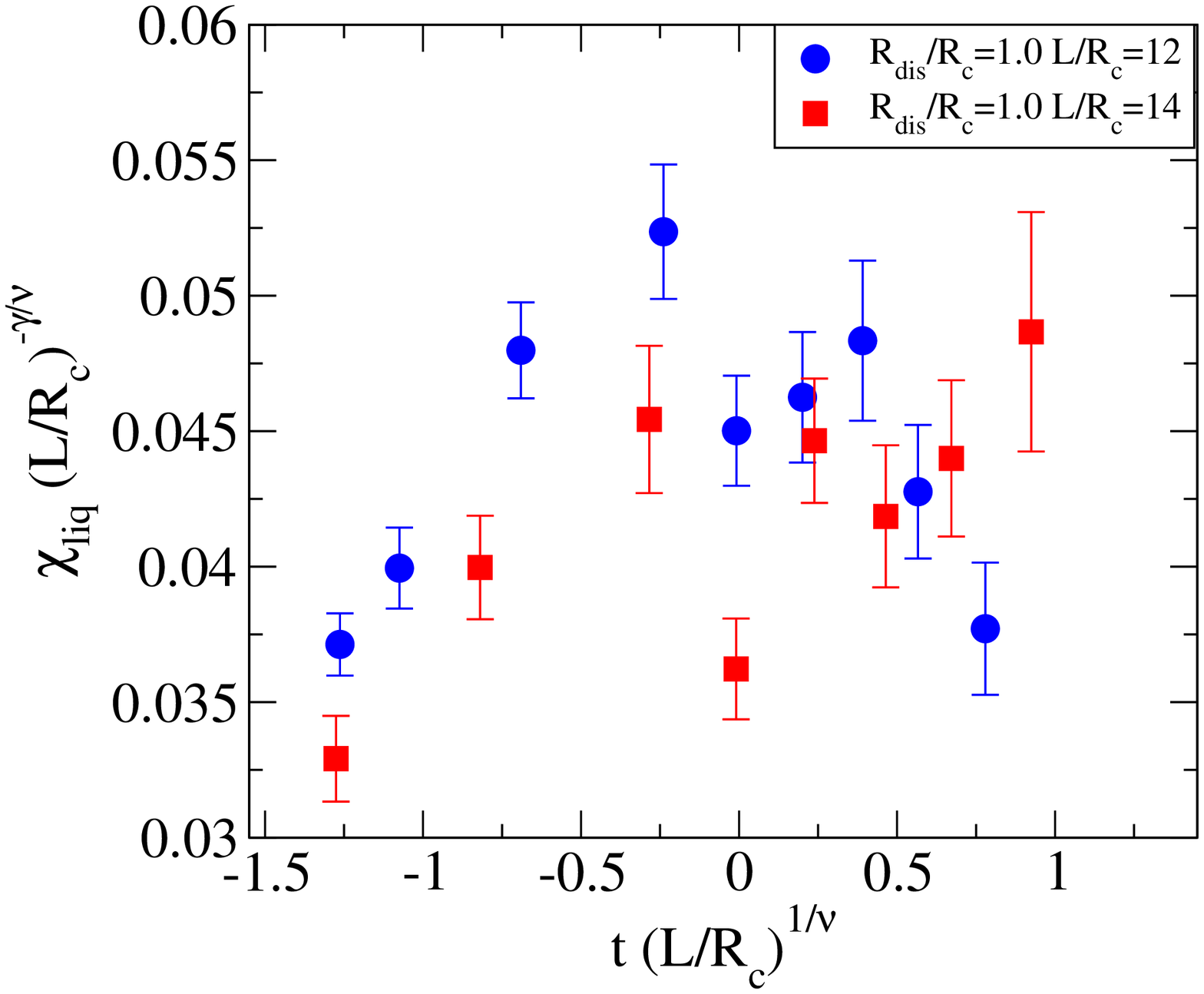} \\
\vspace{0.0cm}
\end{tabular}
\caption{(Color online)
Rescaled connected susceptibilities $(L/R_c)^{-\gamma/\nu} \chi$ as 
a function of $t \, (L/R_c)^{1/\nu}$, where 
$t \equiv \eta^r_p / \eta^r_{p,{\rm crit}} - 1$, for $L/R_c = 12, 14, 16$.
Data 
for $f = 0.4$, $R_{\rm dis}/R_c = 0.2$ (top) and $R_{\rm dis}/R_c = 1$ 
(bottom).
We use \cite{FMY-11} $\gamma/\nu=1.5$, $\nu=0.9$.}
\label{fig:cap6_chi_s}
\end{figure}

As discussed at length in 
Refs.~\cite{deGennes-84,VBL-06,VBL-08,PVCL-08,Vink-09},
the critical transition belongs to the 
RFIM universality class. Some numerical estimates of the RFIM
critical exponents are reported in Table~\ref{tab:RFIM}. It is evident
that there is no general consensus on the estimates and thus we tried all 
different possibilities. Let us first consider the data for 
$R_{\rm dis}/R_c = 0.2$, which are the most precise.
We determine the critical polymer fugacity $z_{p,\rm crit}$,
or equivalently $\eta^r_{p,{\rm crit}}$, by requiring the estimates of
$\chi L^{-\gamma/\nu}$ to collapse onto a single curve as a function 
of $t L^{1/\nu}$, fixing the exponents to the RFIM values. 
If we use the numerical 
values of the exponents employed in Ref.~\cite{VBL-08}, we obtain 
a poor collapse of the data for any choice of the critical polymer 
fugacity.
Apparently, the best collapse is obtained by taking the 
latest numerical estimates of Ref.~\cite{FMY-11}. The rescaled data are 
reported in Fig.~\ref{fig:cap6_chi_s} and allow us to estimate the 
scaling function $F(x)$ appearing in the scaling Ansatz (\ref{FSS}).
This function is universal apart from two rescalings: if $F_1(x)$ and 
$F_2(x)$ are determined for two different transitions belonging to the same 
universality class, then one should have $F_1(x) = a F_2(bx)$, for suitable,
nonuniversal constants $a$ and $b$. We can thus compare the curves appearing in
Fig.~\ref{fig:cap6_chi_s} with 
the analogous ones reported in Ref.~\cite{VBL-08}. Shapes are 
quite similar, although in our case the peak of the scaling functions 
apparently occurs for 
$x\approx 0$, while in Ref.~\cite{VBL-08} the maximum occurs for $x$ 
well below zero. The origin of this difference is unclear: it might 
be due to the different choices for the RFIM critical exponents or to
the neglected scaling corrections.
The results for the critical parameters are reported in 
Table~\ref{tab:cap6_eta_crit}. It is interesting to compare these results
with those which would be obtained by a more naive analysis of the 
diameters. If we compute the intersection of the diameter line 
with an interpolation of the binodal data for $L/R_c = 14$ we would obtain 
$\eta_{p,\rm crit} = 1.02$, $\eta_{c,\rm crit} = 0.20$. The critical 
colloid volume fraction agrees within errors with that reported in 
Table ~\ref{tab:cap6_eta_crit}, while $\eta_{p,\rm crit}$ is slightly
(6\%) underestimated. 

\begin{table}
\begin{center}
\begin{tabular}{cccccc}
\hline
\hline
$f$ & $R_{\rm dis}/R_c$ & $\eta^r_{p,{\rm crit}}$ & $\eta^r_{c,{\rm crit}}$ & 
$\eta_{p,{\rm crit}}$ & $\eta_{c,{\rm crit}}$ \\
\hline
bulk & {} & $1.321(4)$ & $0.575(1)$ & $0.93(2)$ & $0.22(1)$ \\
$0.4$ & $0.2$ & $1.49(6)$ & $0.602(4)$ & $1.08(2)$ & $0.21(1)$ \\
{}    & $1.0$ & $1.67$ & $0.62$ & $1.17$ & $0.22$ \\
\hline
\hline
\end{tabular}
\end{center}
\caption{Estimates of the critical point position.}
\label{tab:cap6_eta_crit}
\end{table}

It is not easy to extend the analysis to the case $R_{\rm dis}/R_c = 1$,
given the very large fluctuations in the data. A reasonable collapse,
see Fig.~\ref{fig:cap6_chi_s}, is obtained for the parameter choices 
given in Table~\ref{tab:cap6_eta_crit}. However, errors cannot be reliably 
estimated. Note that size corrections are quite significant for this
size of the quenched colloids. 
For instance, half of the points for which we observe
bimodal histograms belong to the homogeneous phase in the infinite-volume
limit  (compare the estimate of $\eta_{p,\rm crit}^r$ with the data 
reported in the right-bottom panel of Fig.~\ref{fig:binodal_reser_dis}).

\section{Conclusions} \label{sec5}

\begin{figure}
\begin{center}
\includegraphics[angle=0,width=0.6\textwidth]{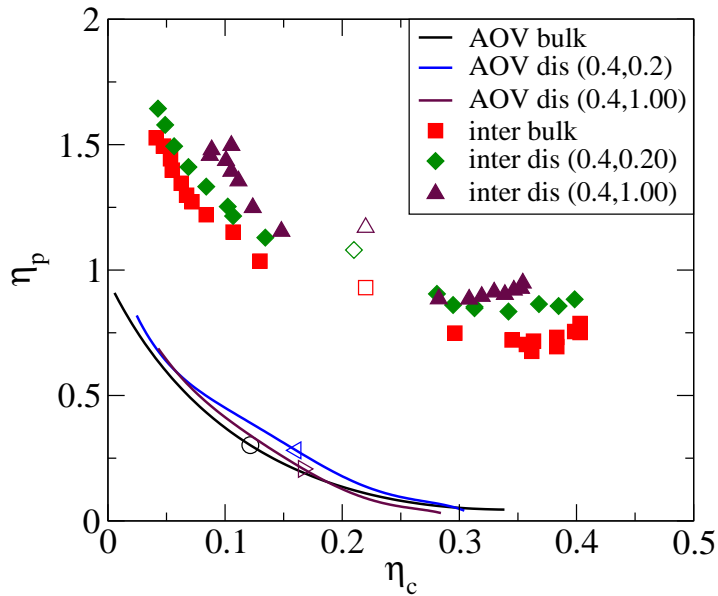}
\end{center}
\caption{(Color online) 
Binodal curves for the interacting-polymer model 
(solid symbols) and for the AOV model (lines). 
Empty symbols give the critical points. }
\label{fig:cap6_bf}
\end{figure}

In this paper we consider a simplified CG model for polymer-colloid 
mixtures. Polymers are represented by point particles interacting by means of 
pair potentials. Models of this type have been shown to be quantitatively
predictive as long as the polymer density is below the 
overlap density, i.e. for $\eta_p\lesssim 1$. Hence, they can be used in the 
study of mixtures in the colloid regime ($q\lesssim 1$), in which
fluid-fluid phase separation occurs for $\eta_p \lesssim 1$. We study 
polymer-colloid segregation in the bulk and in the presence of a porous matrix
for $f = 0.4$ and for two different values of $R_{\rm dis}/R_c$, 0.2 and 1.
In Fig.~\ref{fig:cap6_bf} we summarize our results for the coexistence 
curves and compare them with the AOV ones. The effect of the matrix is
similar in the AOV and in the present interacting model. Disorder 
moves---but the effect is not large---the coexistence curve towards larger 
polymer densities. 
In the colloid-gas phase (for $\eta_c = 0.1$, say),  the observed change 
is quantitatively similar in the two models. The volume fraction
$\eta_p^*$ at coexistence increases by 10-20\% both in the 
interacting and in the AOV case as $f$ increases from $f =0$ (bulk) 
to $f=0.4$. The behavior in the 
colloid-liquid phase is instead quantitatively slightly different. 
In the AOV case, the $f$-dependence of the coexistence curve is very small 
(all curves coincide within errors), while in the interacting 
case the behavior is analogous to that observed in the colloid-gas phase:
$\eta_p^*$ at coexistence increases by 10-20\%. It is important to stress
that this different behavior may be an artifact of the model, since,
as we have already observed several times, binodals cannot be trusted 
for $\eta_c\gtrsim 0.35$.

While binodals do not change significantly with disorder in both cases, 
the critical point position changes significantly. Moreover, the behavior
is quite different in the AOV and in the interacting-polymer model.
In the AOV case the 
critical colloid volume fraction increases significantly as 
$f$ changes from zero (bulk case) to 0.4 and 0.7, while 
$\eta_{p,\rm crit}$ is approximately constant or slightly decreases: 
for $f=0$, $\eta_{c,\rm crit} = 0.1340(2)$ and $\eta_{p,\rm crit} = 0.3562(6)$
\cite{VH-04bis,VHB-05}, while for $f=0.4$ $\eta_{c,\rm crit} \approx 0.19$ 
and $\eta_{p,\rm crit}$ varies between 0.27 and 0.34 depending on 
$R_{\rm dis}/R_c$ \cite{AP-11}.  In the interacting case
instead, the colloid $\eta_{c,\rm crit}$ is approximately independent of 
$f$, while $\eta_{p,\rm crit}$ changes---it increases---with both $f$
and $R_{\rm dis}/R_c$. Note that a similar behavior is observed for the 
$q$ dependence (at least, in the colloid regime) of 
$\eta_{c,\rm crit}$ and $\eta_{p,\rm crit}$ in the bulk \cite{BLH-02}.
In the AOV case, $\eta_{p,\rm crit}$ is essentially independent of $q$,
while $\eta_{c,\rm crit}$ varies significantly with $q$. 
In the interacting-polymer case the opposite occurs: 
$\eta_{p,\rm crit}$ varies with $q$, while $\eta_{c,\rm crit}$ stays 
approximately constant.

In the AOV case we found \cite{AP-11} that colloids could undergo 
capillary condensation: for some values of the parameters, 
a colloid-gas phase in the bulk was in chemical equilibrium with 
a colloid-liquid phase in the matrix. In the present model 
no such phenomenon occurs unless one carefully tunes the system parameters.

The results we have presented rely on simulations of a very simplified model:
first, we use a CG model of the polymers in which each polymer is 
replaced by a monoatomic molecule; second, we use simplified potentials
which allow us to reproduce correctly the thermodynamics in the 
low-density limit, but not the intermolecular distribution functions. 
An improvement of the model is clearly necessary if one wishes to obtain 
quantitatively accurate predictions. A multiblob model \cite{multiblob} 
in which polymers are represented as polyatomic molecules is necessary if one 
wishes to consider larger polymer-to-colloid ratios or to obtain accurate
results for small values of $R_{\rm dis}/R_g$. In the latter case,
one would expect that an accurate modelling requires CG multiblob models in
which the blob size is smaller than $R_{\rm dis}$. 
Studies of the phase behavior of mixtures in random porous matrices 
with these improved models are not feasible with present computer 
resources. However, a less CPU-demanding task would be the study 
of the phase behavior in the presence of {\em regular} arrays of obstacles 
(the quenched colloids could belong to the sites of a regular lattice).
Although less interesting from an experimental point of view (silica gels
have a random distribution of pores) these type of systems could be more 
carefully studied, without relying on many different approximations.
They can thus provide a theoretical laboratory, where one can 
understand the role of quenched obstacles on the phase behavior of these 
systems and thus develop theories that can then be extended to the 
more difficult random case.
Work in this direction is in progress.

\begin{acknowledgments}

It is a pleasure to thank Jean-Pierre Hansen for inspiring comments.
The MC simulations were performed at the INFN Pisa GRID DATA center and on
the INFN cluster CSN4.

\end{acknowledgments}


\begin{thebibliography}{199}

\bibitem{GGRS-99}
L. D. Gelb, K. E. Gubbins, R. Radhakrishnan, and
M. Sliwinska-Bartkowiak,
Rep. Prog. Phys. {\bf 62}, 1573 (1999).


\bibitem{AO-54}
S. Asakura and F. Oosawa, J. Chem. Phys. {\bf 22}, 1255 (1954).

\bibitem{Vrij-76}
A. Vrij, Pure and Appl. Chem. {\bf 48}, 471 (1976).

\bibitem{BLH-02}
P. G. Bolhuis, A. A. Louis, and J. P. Hansen, 
Phys. Rev. Lett. {\bf 89}, 128302 (2002).


\bibitem{GHR-83}
A. P. Gast, C. K. Hall, and W. B. Russell, 
J. Colloid Interface Sci. {\bf 96}, 251 (1983).

\bibitem{LPPSW-92}
H. N. W. Lekkerkerker, W. C. K. Poon, P. N. Pusey, 
A. Stroobants, and P. B. Warren, 
Europhys. Lett. {\bf 20}, 559 (1992).

\bibitem{MF-94}
E. J. Meijer and D. Frenkel, J. Chem. Phys. {\bf 100}, 6873 (1994).

\bibitem{DBE-99}
M. Dijkstra, J. M. Brader, and R. Evans, 
J. Phys.: Condens. Matter {\bf 11}, 10079 (1999).

\bibitem{SLBE-00}
M. Schmidt, H. L\"owen, J. M. Brader, and R. Evans,
Phys. Rev. Lett. {\bf 85}, 1934 (2000);
J. Phys.: Condens. Matter {\bf 14}, 9353 (2002).

\bibitem{DvR-02}
M. Dijkstra and R. van Roij, 
Phys. Rev. Lett. {\bf 89}, 208303 (2002). 

\bibitem{VH-04bis}
R. L. C. Vink and J. Horbach, 
J. Chem. Phys. {\bf 121}, 3253 (2004).

\bibitem{VHB-05}
R. L. C. Vink, J. Horbach, and K. Binder,
Phys. Rev. E {\bf 71}, 011401 (2005).


\bibitem{SSKK-02}
M. Schmidt, E. Sch\"oll-Paschinger, J. K\"ofinger, and G. Kahl,
J. Phys.: Condens. Matter {\bf 14}, 12099 (2002).

\bibitem{VBL-06}
R. L. C. Vink, K. Binder, and H. L\"owen,
Phys. Rev. Lett. {\bf 97}, 230603 (2006).

\bibitem{VBL-08}
R. L. C. Vink, K. Binder, and H. L\"owen,
J. Phys.: Condens. Matter {\bf 20}, 404222 (2008).

\bibitem{PVCL-08}
G. Pellicane, R. L. C. Vink, C. Caccamo, and H. L\"owen, 
J. Phys.: Condens. Matter {\bf 20}, 115101 (2008).

\bibitem{Vink-09}
R. L. C. Vink, Soft Matter {\bf 5}, 4388 (2009).

\bibitem{AP-11}
M. A. Annunziata and A. Pelissetto,
Mol. Phys. {\bf 109}, 2823 (2011).

\bibitem{deGennes-84}
P. G. de Gennes, J. Phys. Chem. {\bf 88}, 6469 (1984).

\bibitem{DSLP-08}
P. G. De Sanctis Lucentini and G. Pellicane,
Phys. Rev. Lett. {\bf 101}, 246101 (2008).

\bibitem{FV-11}
T. Fischer and R. L. C. Vink,
J. Chem. Phys. {\bf 134}, 055106 (2011).

\bibitem{Likos-01}
C. N. Likos, Phys. Rep. {\bf 348}, 267 (2001).

\bibitem{LBHM-00}
A. A. Louis, P. G. Bolhuis, J. P. Hansen, and E. J. Meijer,
Phys. Rev. Lett. {\bf 85}, 2522 (2000).

\bibitem{BLHM-01}
P. G. Bolhuis, A. A. Louis, J. P. Hansen, and E. J. Meijer, 
J. Chem. Phys. {\bf 114}, 4296 (2001).

\bibitem{multiblob}
See, e.g., C. Pierleoni, B. Capone, and J. P. Hansen, 
J. Chem. Phys. {\bf 127}, 171102 (2007),
G. D'Adamo, A. Pelissetto, and C. Pierleoni,
Soft Matter {\bf 8}, 5151 (2012), and references therein.

\bibitem{DH-94}
J. Dautenhahn and C. K. Hall, 
Macromolecules {\bf 27}, 5399 (1994).

\bibitem{BL-02}
P. G. Bolhuis and A. A. Louis, 
Macromolecules {\bf 35}, 1860 (2002).

\bibitem{PH-05}
A. Pelissetto and J.-P. Hansen, 
J. Chem. Phys. {\bf 122}, 134904 (2005).

\bibitem{PH-06}
A. Pelissetto and J. P. Hansen,
Macromolecules {\bf 39}, 9571 (2006).

\bibitem{CMP-06}
S. Caracciolo, B. M. Mognetti, and A. Pelissetto,
J. Chem. Phys. {\bf 125}, 094903 (2006).

\bibitem{Pelissetto-08}
A. Pelissetto, J. Chem. Phys. {\bf 129}, 044901 (2008).

\bibitem{ZVBHV-09}
J. Zausch, P. Virnau, K. Binder, J. Horbach, R. L. C. Vink,
J. Chem. Phys. {\bf 130}, 064906 (2009).

\bibitem{TV-77}
G. M. Torrie and J. P. Valleau, J. Comp. Phys. {\bf 23}, 197 (1977).


\bibitem{MP-92}
E. Marinari and G. Parisi, Europhys. Lett. {\bf 19}, 451 (1992).

\bibitem{LFP-02}
E. Luijten, M. E. Fisher, and A. Z. Panagiotopoulos,
Phys. Rev. Lett. {\bf 88}, 185701 (2002).

\bibitem{Hasenbusch-10}
M. Hasenbusch, Phys. Rev. B {\bf 82}, 174433 (2010).

\bibitem{WB-92}
N. B. Wilding and A. D. Bruce, J. Phys: Condens. Matter {\bf 4}, 3087
(1992).

\bibitem{BW-92}
A. D. Bruce and N. B. Wilding, Phys. Rev. Lett. {\bf 68}, 193 (1992).


\bibitem{BO-97}
I. Bodn\'ar and W. D. Oosterbaan, J. Chem. Phys. {\bf 106}, 7777 (1997).

\bibitem{IOPP-95}
S. M. Ilett, A. Orrock, W. C. K. Poon, and P. N. Pusey, 
Phys. Rev. E {\bf 51}, 1344 (1995).

\bibitem{TdK-99}
R. Tuinier and C. G. de Kruif,
J. Chem. Phys. {\bf 110}, 9296 (1999).

\bibitem{VDDL-96}
N. A. M. Verhaegh, J. S. van Duijneveldt, J. K. G. Dhont, and 
H. N. W. Lekkerkerker, Physica {\bf 230A}, 409 (1996).

\bibitem{TSPEALF-08}
R. Tuinier, P. A. Smith, W. C. K. Poon, S. U. Egelhaaf, D. G. A. L. 
Aarts, H. N. W. Lekkerkerker, and G. J. Fleer,
Europhysics Lett. {\bf 82}, 68002 (2008).

\bibitem{footnoteCP}
References \cite{IOPP-95,TSPEALF-08,BLH-02} do not report 
explicit estimates of the critical points. The numbers reported in 
Table~\ref{tab:cap5_crit}
have been extracted from figures appearing in the original papers.
The reported experimental estimates are obtained from Fig.~2(i)
of Ref.~\cite{IOPP-95} and from Fig.~1 of Ref.~\cite{TSPEALF-08}
(we assume $\eta_{c,\rm crit} = 0.2$ in both cases), the numerical results of 
Ref.~\cite{BLH-02} are taken from their Fig.~3.

\bibitem{FT-08}
G. J. Fleer and R. Tuinier,
Adv. Colloid Interface Sci. {\bf 143}, 1 (2008).

\bibitem{footnote-etarc} 
The values of $z_c$ at coexistence are quite large and correspond, in the 
hard-sphere case, to systems in the solid phase, 
i.e. such that 
$\eta_c^r > \eta_{c,\rm solid} = 0.545(2)$ [W. G. Hoover and F. H. Ree,
J. Chem. Phys. {\bf 49}, 3609 (1968)]. We obtained 
$\eta_c^r$ by solving the equation $z_c R_c^3 = (z_c R_c^3)_{\rm coex} 
f_H(\eta_c^r)$ where $f_H(\eta_c^r)$ was obtained by using 
Hall's equation of state [K. R. Hall, J. Chem. Phys. {\bf 57},
2252 (1972)]. The fugacity $(z_c R_c^3)_{\rm coex}$ 
at the liquid-solid coexistence was computed by using the Carnahan-Starling
equation of state [N. F. Carnahan and K. E. Starling, J. Chem. Phys. {\bf 51}, 
635 (1969), L. L. Lee,  J. Chem. Phys. {\bf 103}, 9388 (1995)]:
$(z_c R_c^3)_{\rm coex} = 3 \eta_{c,\rm liq}/(4 \pi) 
\exp[f(\eta_{c,\rm liq})]$ with
$f(\eta) = \eta (8 - 9\eta + 3\eta^2)/(1 - \eta)^3$ and 
$\eta_{c,\rm liq} = 0.494(2)$.

\bibitem{NB-96}
M. E. J. Newman and G. T. Barkema, 
Phys. Rev. E {\bf 53}, 393 (1996).

\bibitem{HY-01}
A. K. Hartmann and A. P. Young, 
Phys. Rev. B {\bf 64}, 214419 (2001).

\bibitem{MF-02}
A. A. Middleton and D. S. Fisher, 
Phys. Rev. B {\bf 65}, 134411 (2002).

\bibitem{FMY-11}
L. A. Fernandez, V. Mart\'\i n-Mayor, and D. Yllanes,
Phys. Rev. B {\bf 84}, 100408 (2011).


\end{thebibliography}
\end{document}